\documentclass{aa}

\makeatletter
\renewcommand*\aa@pageof{, page \thepage{} of \pageref*{LastPage}}
\makeatother

\usepackage[T1]{fontenc}
\usepackage{ae,aecompl}
\usepackage{natbib,twoopt}
\usepackage{booktabs}

\usepackage{ulem}

\usepackage{graphicx}   
\usepackage{amsmath}    
\usepackage{amssymb}    
\usepackage{gensymb}    
\usepackage{multirow}
\usepackage{mathtools}
\usepackage{IEEEtrantools}
\usepackage{float}
\usepackage{rotating}
\usepackage{bm}
\usepackage{adjustbox}             
\usepackage{subcaption}        
\usepackage{threeparttable}
\usepackage{txfonts} 
\usepackage{footmisc}
\usepackage{hyperref}
\hypersetup{
    colorlinks=true,
    linkcolor=blue,
    filecolor=blue,      
    urlcolor=blue,
    citecolor=blue,
}
\usepackage{pdflscape}

\usepackage{scalefnt}
\usepackage{xcolor}                 

\LetLtxMacro{\oldtextsc}{\textsc}
\renewcommand{\textsc}[1]{\oldtextsc{\scalefont{1.2}#1}}

\usepackage{relsize}

\newcommand{\obs}{\hbox{$\mathrm{_{obs}}$}}
\newcommand{\keV}{\mathrm{keV}}

\newcommand{\phcm}{\hbox{$\mathrm{ph\,cm}^{-2}$}}
\newcommand{\phscm}{\hbox{$\mathrm{ph\,s^{-1}\,cm^{-2}}$}}

\newcommand{\phscmkeV}{\hbox{$\mathrm{ph\,s^{-1}\,cm^{-2}\,keV^{-1}}$}}

\newcommand{\ergs}{\hbox{$\mathrm{erg\,s}^{-1}$}}
\newcommand{\ergcm}{\hbox{$\mathrm{erg\,cm}^{-2}$}}
\newcommand{\ergskeV}{\hbox{$\mathrm{erg\,s^{-1}\,keV^{-1}}$}}
\newcommand{\ergscm}{\hbox{$\mathrm{erg\,s^{-1}\,cm^{-2}}$}}

\newcommand{\yrMpc}{\hbox{$\mathrm{yr^{-1}\,Mpc^{-3}}$}}
\newcommand{\MsunyrMpc}{\hbox{$\mathrm{M_{\rm{\odot}}\,yr^{-1}\,Mpc^{-3}}$}}
\newcommand{\yr}{\hbox{$\mathrm{yr^{-1}}$}}

\newcommand{\yrGpc}{\hbox{$\mathrm{yr^{-1}\,Gpc^{-3}}$}}

\newcommand{\Msun}{\hbox{M$_{\rm{\odot}}$}}

\newcommand{\dd}{\mathrm{d}}

\newcommand{\ncoll}{\hbox{$\dot{n}_\mathrm{cc}$}}
\newcommand{\nGRB}{\hbox{$\dot{n}_\mathrm{GRB}$}}
\newcommand{\nGRBo}{\hbox{$\dot{n}^0_\mathrm{GRB}$}}
\newcommand{\NGRB}{\hbox{$N_{\rm GRB}$}}
\newcommand{\Ep}{\hbox{$E_p$}}
\newcommand{\EpL}{\hbox{$E_p - L$}\xspace}
\newcommand{\Eiso}{\hbox{$E_{\rm iso}$}}
\newcommand{\sigmaEp}{\hbox{$\mathrm{\sigma_{E_p}}$}}
\newcommand{\Epo}{\hbox{$E_{p0}$}}

\newcommand{\Eminobs}{\hbox{$E_\mathrm{min,obs}$}}
\newcommand{\Emaxobs}{\hbox{$E_\mathrm{max,obs}$}}
\newcommand{\dl}{\hbox{$D_L$}}

\newcommand{\Liso}{\hbox{$L_{\rm iso}$}}
\newcommand{\Lmin}{\hbox{$L_{\rm min}$}}
\newcommand{\Lmax}{\hbox{$L_{\rm max}$}}

\newcommand{\Lstar}{\hbox{$L_*$}}
\newcommand{\alphaA}{\hbox{$\alpha_{\rm A}$}}
\newcommand{\kevol}{\hbox{$k_{\rm evol}$}}
\newcommand{\Cvar}{\hbox{$C_{\rm var}$}}
\newcommand{\Tnt}{\hbox{$T_{90}$}}
\newcommand{\Av}{\hbox{A$_V$}}
\newcommand{\Tsim}{\hbox{$T_{\rm sim}$}}
\newcommand{\Npk}{\hbox{$N^{\rm pk}$}}
\newcommand{\Chisq}{\hbox{$\mathrm{\chi^2}$}}
\newcommand{\lnL}{\hbox{$\mathrm{\ln\mathcal{L}}$}}

\newcommand{\Tlive}[1]{\hbox{$T^{\mathrm{#1}}_{\mathrm{live}}$}}
\newcommand{\mOmega}[1]{\hbox{$\langle\,\Omega^{\mathrm{#1}}\,\rangle$}}

\newcommand{\Npksub}[3]{\hbox{$N^{\rm pk}_{\rm {#1}-{#2}\,{#3}}$}}
\newcommand{\Nb}[1]{\hbox{$N_{\rm {#1}}$}}
\newcommand{\Rate}[2]{\hbox{$R^{\rm {#1}}_{\rm #2}$}}

\newcommand{\BATSE}{\textit{CGRO}/BATSE}
\newcommand{\GBM}{\textit{Fermi}/GBM}
\newcommand{\Swift}{\textit{Swift}/BAT}

\newcommand{\eBATsix}{\hbox{\textit{Swift}/eBAT6}}
\newcommand{\bGBM}{\hbox{\GBM$^{bright}$}}

\begin{document}

\title{Constraining the intrinsic population of Long Gamma-Ray Bursts: implications for spectral correlations, cosmic evolution and their use as tracers of star formation}
\author{
J. T. Palmerio$^{1,2}$\thanks{E-mail: palmerio@iap.fr}
\and F. Daigne$^{1}$
}
\institute{
Sorbonne Universit\'e, CNRS, UMR7095, Institut d'Astrophysique de Paris, F-75014, Paris, France
\and
GEPI, Observatoire de Paris, PSL University, CNRS, 5 Place Jules
Janssen, F-92190 Meudon, France
}

\date{Accepted XXX. Received YYY; in original form ZZZ}

\label{firstpage}

\abstract{}
{Long Gamma-Ray Bursts (LGRBs) have been shown to be powerful probes of the Universe, in particular to study the star formation rate up to very high redshift ($z\sim9$).
Since LGRBs are produced by only a small fraction of massive stars, it is paramount to have a good understanding of their underlying intrinsic population in order to use them as cosmological probes without introducing any unwanted bias.
The goal of this work is to constrain and characterise this intrinsic population.
}
{We developed a Monte Carlo model where each burst is described by its redshift and its properties at the peak of the lightcurve.
We derived the best fit parameters by comparing our synthetic populations to carefully selected observational constraints based on the \BATSE, \GBM\ and \Swift\ samples with appropriate flux thresholds.
We explored different scenarios in terms of cosmic evolution of the luminosity function and/or of the redshift distribution as well as including or not the presence of intrinsic spectral-energetics (\EpL) correlations.
}
{We find that the existence of an intrinsic \EpL\ correlation is preferred but with a shallower slope than observed ($\alphaA\sim0.3$) and a larger scatter ($\sim0.4$~dex).
We find a strong degeneracy between the cosmic evolution of the luminosity and of the LGRB rate, and show that a sample both larger and deeper than SHOALS by a factor of three is needed to lift this degeneracy.
}
{
The observed \EpL correlation cannot be explained only by selection effects although these do play a role in shaping the observed relation.
The degeneracy between cosmic evolution of the luminosity function and of the redshift distribution of LGRBs should be included in the uncertainties of star formation rate estimates; these amount to a factor of 10 at $z=6$ and up to a factor of 50 at $z=9$.
}

\authorrunning{J. T. Palmerio \& F. Daigne}
\titlerunning{Constraining the intrinsic population of LGRBs}

\maketitle

\section{Introduction}\label{sec:intro}
Gamma-Ray Bursts (GRBs) are the most powerful electromagnetic phenomena, associated to the relativistic ejection following the birth of a stellar mass compact object (black hole or magnetar, see e.g. \citealt{Piran2005,Kumar2015}).
GRBs exhibit two main radiative phases:
(i) the prompt emission which is commonly detected in the hard X-rays and soft $\gamma$-rays (1~keV - 10~MeV) and usually lasts a few hundreds of milliseconds to a few hundreds of seconds;
(ii) the afterglow emission, which is initially bright, decays rapidly, and peaks successively in X-rays, optical and radio
\citep[see e.g.][]{Gehrels2009,Gehrels2013}.
GRBs have proven themselves to be powerful probes of the Universe owing to some unique properties.
They form up to very high redshift (spectroscopically confirmed at $z\sim8.2$, \citealt{Salvaterra2009,Tanvir2009} and using photometry only at $z\sim9$, \citealt{Cucchiara2011}) and can be detected thanks to their $\gamma$-ray light which is largely unaffected by dust and hydrogen absorption.
The transient, fading nature of their afterglow benefits from time-dilation due to cosmological expansion \citep{Lamb2000}; 1 day after the prompt emission on Earth corresponds to 6 hours in the source frame at $z = 3$ and 2 hours at $z = 10$, essentially catching the afterglow earlier in its light curve and thus brighter as redshift increases.

Among the various classes of GRBs, the case of \textit{long} GRBs (LGRBs, which have a duration of the prompt emission longer than $\sim 2$~s, \citealt{Mazets1981,Kou1993}) is the most promising for the study of the distant Universe.
These are the most frequent GRBs and both theoretical progenitor models ("collapsar" model, \citealt{Woosley1993,Paczynski1998}) and observations have firmly associated them with the collapse of certain massive stars.
Indeed, they generally occur in faint, blue, low-mass star-forming galaxies \citep{Lefloch2003,Savaglio2009,Palmerio2019}, and in the UV-brightest regions of their hosts \citep{Fruchter2006,Svensson2010,Lyman2017}.
In addition, the majority of low-redshift LGRBs are found in association with core-collapse supernovae \citep{Bloom1999,Hjorth2003,Kruehler2017,Izzo2017,Melandri2019,Izzo2020}.
Due to the short-lived nature of their massive star progenitors, LGRBs are expected to occur up to very high redshift, possibly in association with the first generation of stars \citep{Bromm2006}, and thus can be used as lighthouses to study galaxies \citep[e.g.][]{Lefloch2006,Perley2013,Vergani2017} and the intergalactic medium at high redshift \citep[e.g.][]{Fynbo2006a,Hartoog2015,Selsing2019,Tanvir2019,Vielfaure2020}.

From their association with certain massive stars, LGRBs as a population are considered potential tracers of star formation \citep[e.g.][]{Lamb2000,Porciani2001,Kistler2008,Robertson2012,Vangioni2015}.
However, the precise link between star formation and LGRBs remains shrouded in uncertainty for two main reasons.
The first reason is that this link depends on many factors which are sometimes poorly constrained and which may evolve with redshift such as the stellar Initial Mass Function (IMF), the properties of the progenitor star (e.g. mass range, metallicity distribution, initial rotation distribution), the fraction of binary progenitors \citep[see e.g.][]{Chrimes2020} or the LGRB luminosity, spectrum and jet opening angle.
The second reason is the difficulty in obtaining large, unbiased and redshift-complete observational samples of LGRBs as they require a rapid response from ground follow-up and significant telescope time to get meaningful statistics.
In practice we often have to chose between either biased or incomplete or smaller samples \citep[see e.g.][ for the case of smaller but unbiased and highly complete samples]{Hjorth2012,Salvaterra2012,Kruhler2015,Perley2016b}.

The goal of population models is to overcome the limitations of biased or incomplete samples by modelling the underlying intrinsic population and fitting it to carefully selected observational samples.
This allows one to then constrain characteristics of the intrinsic population such as the redshift distribution, the rate, the evolution of the energetics with redshift or the correlations between physical parameters.
Over the last 20 years, population models have become an effective approach to study GRBs due to growing sample sizes and increasingly affordable computing power.
\citet{Porciani2001} studied the luminosity function of LGRBs using the peak flux distribution of \BATSE\ from \citet{Kommers2000} and estimated the rate of LGRBs under the hypothesis that their redshift distribution follows the cosmic star formation rate density (CSFRD).
They used a fixed Band function \citep[][see App.~\ref{app:band}]{Band1993} with parameters $\alpha=-1$, $\beta=-2.25$, and $\Ep=511~\keV$ for the LGRB spectrum.
Assuming a constant LGRB production efficiency from stars, they found $(1-2) \times 10^{-6}$ LGRBs pointing towards us per core-collapse.
\citet{Daigne2006} used a similar approach but, in addition to including observations of the peak energy distribution of \BATSE\ LGRBs, they allowed for three scenarios regarding the redshift distribution of LGRBs.
These scenarios correspond to three different hypotheses about the LGRB production efficiency: constant, mildly increasing and strongly increasing with redshift.
They also used a Band spectrum but added realistic distributions for the values of $\alpha$ and $\beta$ as well as two different scenarios for the source-frame peak energy distribution: log-Normal or correlated to the peak luminosity.
Using the observed redshift distribution from early \textit{Swift} results of \citet{Jakobsson2006} as a cross-check, they found evidence for an increasing LGRB production efficiency with redshift.
With the advent of \textit{Swift} and thanks to dedicated follow-up campaigns, samples of GRBs with a significant fraction of redshift measurements were obtained, allowing for two new approaches to emerge.
The first approach is to model the redshift recovery efficiency as in \citet{Wanderman2010}.
These authors directly inverted the observed luminosity-redshift distribution of a sample of 120 \textit{Swift} LGRBs, assuming a fixed Band spectrum as in \citet{Porciani2001}, to obtain the intrinsic luminosity function and redshift distribution.
They found some evidence that the LGRB rate does not follow the CSFRD, in agreement with \citet{Daigne2006}.
Using a different approach, \citet{Salvaterra2012} designed a complete and unbiased sample of 58 bright \textit{Swift} LGRBs named BAT6 (later extended to 99 LGRBs by \citealt{Pescalli2016}), essentially paying the cost of sample size in order to achieve high redshift completeness and avoid the uncertainties caused by modelling the redshift recovery efficiency.
They found evidence for strong redshift evolution of the luminosity function although they noted the degeneracy with the redshift evolution of the LGRB efficiency and suggested both might be at play.

Building on these works, we present in this paper our own population model which combines the latest, largest and most diverse datasets of LGRBs available to date with currently relevant scenarios in order to investigate some of the questions raised above.
We fit our population model parameters using an intensity constraint based on a large sample of \BATSE\ LGRBs including faint events, a spectral constraint based on a sample of bright \GBM\ LGRBs with well measured spectral parameters, and a redshift constraint based on the extended BAT6 sample.
Finally, we discuss the implications of our results in terms of intrinsic spectral-energetics correlations, of soft bursts, of cosmic evolution of the LGRB luminosity and/or rate, of the use of LGRBs as tracers of star formation at high redshift and of strategies for detecting high redshift bursts.

The paper is organised as follows.
The assumptions and scenarios we use to model the intrinsic LGRB population are presented in Sect.~\ref{sec:LGRB_pop_mod} while the observations we used to constrain our model are presented in Sect.~\ref{sec:obs_constraints}.
Results for the best fitting scenarios are given in Sect.~\ref{sec:results}, a discussion on the implications and predictions of our model is presented in Sect.~\ref{sec:disc}, and our conclusions are compiled in Sect.~\ref{sec:concl}.
Finally, additional relevant material including the statistical framework used and details about our method are provided in the Appendix.
All errors are reported at the 1$\sigma$ confidence level unless stated otherwise.
We use a standard cosmology: $\Omega_{\rm m} = 0.27$, $\Omega_{\Lambda} = 0.73$, and $H_{0} = 71$~km~s$^{-1}$~Mpc$^{-1}$.

\section{Modelling the intrinsic LGRB population}\label{sec:LGRB_pop_mod}
\renewcommand{\arraystretch}{1.2}
\begin{table*}
\begin{center}
\caption{\textbf{Summary of the properties and calculated quantities used to describe the LGRBs in our population model.}
In each table, the Pk type refers to quantities defined at peak brightness of the LGRB while the Ti signifies the quantity is time-integrated (i.e. it depends on the lightcurve of the LGRB).
The first table lists the properties describing a single LGRB, always defined in the source frame.
The top three properties' distributions are adjusted by comparing the population to the observational constraints described in Sect.~\ref{sec:obs_constraints}.
They are completed by the next two properties, with fixed distribution (see Sect.~\ref{subsec:pop_spectr}) to entirely describe the emission at the peak brightness.
The last two properties are only used for additional cross-checks described in Sect.~\ref{subsec:cross_check}.
Their distributions are computed (see Sect.~\ref{subsec:pop_Cvar}).
The second table lists the main quantities that are calculated from the physical properties in the first table, usually in the observer frame.
They are used to build the mock samples as described in Sect.~\ref{subsec:pop_mock_samples}.\\
}
\label{tab:pop_quant}
\noindent\textbf{Properties describing an LGRB:}
\begin{adjustbox}{max width=\textwidth}
\centering
\begin{tabular}{lccll}
\toprule
Property
& Units & Type & Distribution & Description                                                                   \\ 
\toprule
$z$      & -     & -    & Adjusted     & The redshift of the LGRB source                                                \\
$L$      & \ergs & Pk   & Adjusted     & The isotropic-equivalent bolometric peak luminosity                                \\
\Ep      & keV   & Pk   & Adjusted     & The peak energy of the $L_E/E$ photon spectrum in the source frame                \\
\cmidrule(lr){1-5}
$\alpha$ & -     & Pk   & Fixed        & The low-energy slope of the photon spectrum $L_E/E$ in the source frame\\
$\beta$  & -     & Pk   & Fixed        & The high-energy slope of the photon spectrum $L_E/E$ in the source frame\\
\cmidrule(lr){1-5}
\Tnt     & s     & Ti   & Computed     & The duration over which 5 to 95 \% of photons are emitted in the source frame  \\
\Cvar    & -     & Ti   & Computed     & The variability coefficient defined as the mean luminosity divided by the peak luminosity \\
\bottomrule
\vspace{0pt}
\end{tabular}
\end{adjustbox}
\noindent\textbf{Calculated quantities:}
\begin{adjustbox}{max width=\textwidth}
\centering
\begin{tabular}{lcll}
\toprule
Quantity & Units  & Type  & Description   \\ 
\toprule
\Ep\obs  & keV    & Pk    & The peak energy of the $E^2\,N_{E_\mathrm{obs}}$ spectrum in the observer frame, $\Ep\obs = \Ep/(1+z)$.              \\
\Npk     & \phscm & Pk    & The peak photon flux, calculated from $z$, $L$, \Ep, $\alpha$, $\beta$ for any $\left[E_\mathrm{min,obs};E_\mathrm{max,obs}\right]$ (see Eq.~\ref{eq:pflx}) \\
$F^{\rm pk}$ & \ergscm & Pk    & The peak energy flux, calculated from $z$, $L$, \Ep, $\alpha$, $\beta$ for any $\left[E_\mathrm{min,obs};E_\mathrm{max,obs}\right]$ (see Eq.~\ref{eq:erg_pflx}) \\
\cmidrule(lr){1-4}
\Tnt\obs & s     & Ti     & The duration over which 5 to 95 \% of photons are received in the observer frame, $\Tnt\obs = \Tnt\,(1+z)$ \\
$\mathcal{N}$ & \phcm & Ti & The photon fluence, calculated from $z$, $L$, \Ep, $\alpha$, $\beta$, \Tnt\ and \Cvar\ for any $\left[E_\mathrm{min,obs};E_\mathrm{max,obs}\right]$ (see Eq.~\ref{eq:flnc}) \\
$\mathcal{F}$ & \ergcm & Ti & The energy fluence, calculated from $z$, $L$, \Ep, $\alpha$, $\beta$, \Tnt\ and \Cvar\ for any $\left[E_\mathrm{min,obs};E_\mathrm{max,obs}\right]$ (see Eq.~\ref{eq:erg_flnc}) \\
\Eiso    & erg   & Ti      & The isotropic-equivalent bolometric energy, $\Eiso = \Cvar\,\Tnt\,L$                                 \\
\bottomrule
\end{tabular}
\end{adjustbox}
\end{center}

\end{table*}
\renewcommand{\arraystretch}{1.}

We should specify to lift any ambiguity that throughout the rest of this paper the term {\it intrinsic} is used to qualify the entire, true LGRB population, without any selection criteria (e.g. {\it intrinsic} redshift distribution versus redshift distribution of the \textit{Swift} sample)
while the term {\it source-frame} will be reserved to refer to properties as measured without the effect of cosmological redshift dilation (e.g. {\it source-frame} \Ep, versus {\it observed} \Ep, noted \Ep\obs; when not specified the default is source-frame).

Each burst in our model is described by the following properties: a redshift, a peak luminosity and a spectrum.
For the majority of the work presented in this paper, these properties alone are enough to compute all the observed quantities we are interested in, and in particular the peak flux. 
However, for some specific studies described in Sect.~\ref{subsec:cross_check}, other observed quantities such as duration or fluence are necessary.
For this purpose, we add two properties to fully describe an LGRB: a source-frame duration and a variability coefficient which corresponds to the ratio of the mean luminosity (averaged over the duration of the burst) over the peak luminosity.
One intrinsic population is thus described by the probability density function of all these physical properties.
This allows us to generate a synthetic population with a large number of bursts, \NGRB, by Monte Carlo sampling of these intrinsic distributions.
We then aim to constrain the parameters of these distributions by carefully comparing the generated LGRB population to real observed samples of LGRBs.
For this purpose we generate mock samples for different instruments, based on simple selection criteria (e.g. peak flux threshold significantly above the detector threshold) to avoid any modelling of a complex instrumental efficiency, and compare to the corresponding real observed sample.
Since our LGRB population is generated by Monte Carlo sampling, its accuracy depends on \NGRB; we tested different values and settled on $\NGRB=10^6$ as a compromise between population accuracy and computational time.
\subsection{Properties and quantities describing one LGRB}
More specifically, each LGRB in our model is described by a redshift $z$, four peak properties and two additional time-integrated properties which are used only for some cross-checks in Sect.~\ref{subsec:cross_check}.
The four peak properties are defined as occurring when the LGRB is at peak brightness and are independent of the LGRB duration; they are:
\begin{itemize}
    \renewcommand\labelitemi{--}
    \item $L$, the isotropic-equivalent bolometric peak luminosity in units of \ergs, defined as \mbox{$L =\int_0^\infty  L_E\, \dd E$}, where $L_E$ is the source-frame power density in units of [\ergskeV].
    \item $E_p$, the energy at which the $E\,L_E$ spectrum peaks in the source-frame, in units of keV.
    \item $\alpha$, the low-energy slope of the photon spectrum $L_E/E$.
    \item $\beta$, the high-energy slope of the photon spectrum $L_E/E$.
    \item In addition to these four quantities, the intrinsic  shape of the photon spectrum $L_E/E$ is assumed to be given by a Band function \citep{Band1993}.
\end{itemize}

The peak brightness of an LGRB depends on the timescale with which its lightcurve is sampled; in the following, unless stated otherwise, this timescale is assumed to be 1.024\,s in the observer frame (the typical timescale for measuring peak fluxes).
This assumption is ubiquitous for these types of studies, but in reality it introduces a small error on the peak flux estimations because of time dilation; 1.024 s in the observer frame corresponds to different timescales in the source frame depending on the redshift.
However this does not impact significantly the calculated values of the peak flux; a rough estimate puts this correction between 1, 7, 13, 18\% at redshifts 0.1, 1, 3, 6 respectively \citep{HeussaffThesis}.

The two time-integrated properties which depend on the lightcurve of the LGRB are:
\begin{itemize}
    \renewcommand\labelitemi{--}
    \item $T_{90}$, the duration over which 5 to 95\% of photons are emitted in the source frame.
    \item $C_{\rm var}$, the variability coefficient defined as the mean luminosity $\bar{L}$ divided by the peak luminosity $L$.
\end{itemize}

Using these aforementioned properties, it is possible to calculate the peak photon flux \Npk\ for any instrument observing between \Eminobs\ and \Emaxobs\ (in units of [\phscm]) with:
\begin{align}\label{eq:pflx}
    \Npk & = \int_{\Eminobs}^{\Emaxobs} N_{E_\mathrm{obs}}(E\obs)\,\dd E\obs\\
    & = \frac{(1+z)}{4\pi\,D^2_\mathrm{L}(z)}\int_{\Eminobs\,(1+z)}^{\Emaxobs\,(1+z)} \frac{L_E}{E} \, \dd E
\end{align}
where $N_{E_\mathrm{obs}}$ is the observed photon flux density (in units of [\phscmkeV]) and \dl\ is the luminosity distance.

Similarly we can calculate the peak energy flux $F^{\rm pk}$ for any instrument observing between \Eminobs\ and \Emaxobs\ (in units of [\ergscm]) with:
\begin{equation}\label{eq:erg_pflx}
F^{\rm pk}=\frac{1}{4\pi\,D^2_\mathrm{L}(z)}\int_{\Eminobs\,(1+z)}^{\Emaxobs\,(1+z)}L_E\,\dd E
\end{equation}

A summary of all the properties and calculated quantities is presented in Tab.~\ref{tab:pop_quant}.
The paragraphs below are devoted to presenting in detail the assumed probability density distributions for each property, and the parameters of these distributions that we want to constrain from observations.

\subsection{Intrinsic redshift distribution}\label{subsec:pop_zdistr}
The intrinsic redshift distribution is perhaps one of the most important ingredients of an LGRB population model because
any constraint on this distribution has critical implications for the  identification of LGRB progenitors and the use of LGRBs as tracers of star formation.
However, it remains poorly constrained due to the small fraction of GRBs with measured redshifts ($\sim$ 30 \% for {\it Swift} GRBs) leading to significant uncertainties regarding its shape; for these reasons many authors have used various functional forms to represent it \citep[see e.g.][]{Porciani2001,Daigne2006,Wanderman2010,Salvaterra2012,AmaralRogers2016}.

In this work we chose to use a simple functional form, with only 3 free parameters, which could adequately represent the cosmic star formation rate density (CSFRD) while also having the freedom to deviate from it at high or low redshift independently.
The comoving rate density of LGRBs \nGRB\ is thus parametrised in our model as:
$$
\dot{n}_\mathrm{GRB}(z) = \nGRBo\, f(z;a,b,z_m)~~~~~~~~~~~~~~~~~~~[\yrMpc]
$$
with $f(z)$ a broken exponential:
\begin{equation}\label{eq:z_distr}
f(z;a,b,z_m)=\begin{cases}
e^{az} & z < z_m\\
e^{bz}\, e^{(a-b)z_m} & z \geq z_m
\end{cases}\, ,
\end{equation}
where $z_m$ is the redshift of the break, and $a$ and $b$ are the low- and high-redshift slopes respectively and \nGRBo\ is a normalisation given by our model (see Sect.~\ref{subsec:pop_norm}).

We can relate \nGRB\ to the cosmic star formation rate density by introducing the LGRB efficiency $\eta(z)$ defined as $$ \nGRB(z) = \eta(z)\,\ncoll(z)$$ where $\ncoll$ is the comoving rate density of core-collapses (in units of [\yrMpc]).
We then assume that $\ncoll$ is related to the cosmic star formation rate density by: 
$$ \ncoll(z) = \frac{p_{\rm cc}(z)}{\bar{m}(z)}\,\dot{\rho}_*(z)$$
where $\dot{\rho}_*(z)$ is the cosmic star formation rate density in units of [\MsunyrMpc] and $\bar{m}$ is the mean mass deduced from the stellar initial mass function (IMF): 
$$ \bar{m}(z) = \int_{m_{\rm inf}}^{m_{\rm sup}}\,m\,I(m,z)\,\dd m~~~~~~~~~~[\Msun]$$
$p_{\rm cc}(z)$ is the fraction of formed stars ending with a core-collapse, given by:
$$ p_{\rm cc}(z) = \int_{m_{cc}}^{m_{\rm sup}}\,I(m,z)\,\dd m $$
where $m_{cc}$ is the minimum mass at which a core-collapse can occur.
In our case we used a Salpeter \citep{Salpeter1955} stellar IMF with a slope of 1.35 for $I(m,z)$, $m_{cc}=8\,\Msun$, $m_{\rm inf}=0.1\,\Msun$, and $m_{\rm sup}=100\,\Msun$.
Using the numeric values quoted previously, this finally leads to:
\begin{equation}\label{eq:ndot_GRB}
\nGRB(z) =  \,\frac{\eta(z) \, \dot{\rho}_*(z)}{135\, \Msun}
\end{equation}
In the following, we assume $p_{cc}$ and $\bar{m}$ to be constant with redshift, meaning any discrepancy between the redshift distribution of LGRBs and the cosmic star formation rate density is caused by the redshift evolution of the LGRB efficiency $\eta(z)$.
As discussed in Sect.~\ref{subsec:disc_LGRB_SFR} such an evolution should be directly linked to the physics of LGRB progenitors.

The cosmic star formation rate density is then taken to have a similar shape as the  broken exponential defined above:
$$
\dot{\rho}_*(z)  = 
\dot{\rho}^0_*\, f(z;a_*,b_*,z_{m,*})~~~~~~~~~~~~~[\Msun\,\yrMpc]
$$
with 
$\dot{\rho}^0_*=2.8\times10^{-2}$\,\Msun\,\yrMpc,
$a_*=1.1$, $b_*=-0.57$ and $z_{m,*}=1.9$; these values are obtained by fitting the broken exponential form of Eq.~\ref{eq:z_distr} to the slightly more complex functional form of \citet{Springel2003} with the parameter values of model 3 from \citet{Vangioni2015} which are deduced of the observed cosmic formation rate extracted by \citep{Behroozi2013} from galaxy luminosity functions measurements including high redshift measurements at $z\simeq 8-10$ \citep{Oesch2014,Oesch2015,Bouwens2015}, and are compatible with the chemical enrichment history of the Universe and with the constraints on reionisation from the Cosmic Microwave Background data (see \citealt{Vangioni2015} for details).

The particular case in which $\eta$ is constant with redshift corresponds to $a=a_*$, $b=b_*$ and $z_m=z_{m,*}$. 
Then the efficiency $\eta$ is equal to 
$$
\eta_0 = 135\,\Msun \frac{\nGRBo}{\dot{\rho}^0_*}\,=\frac{\nGRBo}{2.1\times10^{-4}\, \yrMpc}\, .
$$

In the more general case where $a$, $b$, and $z_m$ are different from the cosmic star formation rate density values $a_*$, $b_*$ and $z_{m,*}$, we can obtain the evolving LGRB efficiency $\eta(z)$ by rearranging Eq.~\ref{eq:ndot_GRB}:
\begin{equation}\label{eq:LGRB_eff}
\eta(z) = \eta_0 \,\frac{f(z;a,b,z_m)}{f(z;a_*,b_*,z_{m,*})}    
\end{equation}

\subsection{Intrinsic luminosity function}\label{subsec:pop_LF}
The luminosity function is another widely studied distribution of LGRB populations with a variety of different functional forms used \citep[e.g.][]{Salvaterra2012,Wanderman2010,Pescalli2016}.
In our model, due to its ubiquitous use throughout astronomy and its flexibility, we described it using a Schechter function \citep{Schechter1976} as:
\begin{equation}
\phi(L) = A
\begin{cases} 
\left(\frac{L}{L_*}\right)^{-p}\times \exp\left(-\frac{L}{L_*}\right) & L > L_\mathrm{min}\\
0 & L \leq L_\mathrm{min}
\end{cases}
\end{equation}
where $A$ is a normalization factor given by $1 = \int_{0}^{\infty} \phi(L)\,\dd L$. 
In practice this means we are actually working with a probability density but we will continue refering to it as a luminosity function out of convenience.
This functional form has the advantage of having only 3 parameters: the minimum luminosity \Lmin, the break luminosity \Lstar\ and the slope $p$. 
It is worth noting that this form is very similar to a simple power law with the difference that the break at the high-luminosity end is smooth rather than sudden which is slightly more realistic (see the discussion in \citealt{Atteia2017}).
We also tested luminosity models with a broken power law but found little success in deriving meaningful constraints on all parameters simultaneously\footnote{When the high-luminosity slope is well constrained, $L_b$ is close to \Lmin\ and the low-luminosity slope is poorly constrained; conversely when the low-luminosity slope is well constrained, $L_b$ is close to \Lmax\ and the high-luminosity slope is poorly constrained.}.
For this reason, and in order to have a lower number of free parameters, we chose to use the Schechter function.
Furthermore, after some exploration we decided to fix \Lmin\ since it cannot be constrained from current observations as it requires seeing the turnover at low peak fluxes in the $\log{N}$-$\log{P}$ diagram, which is to date unobserved \citep{Stern2001}.
The value chosen was $\Lmin=5\times10^{49}\,$\ergs, which corresponds to the lowest luminosity burst in the \eBATsix\ sample (see Sect.~\ref{subsec:disc_XRFs} for a discussion on the low-luminosity population in our model); this value is similar to the ones assumed by other studies (typically taken between $10^{48}$ and $10^{50}\,$\ergs).
It should be noted that this parameter strongly affects the normalisation of our model and in particular the global LGRB rate (which we will discuss in Sect.~\ref{sec:disc}), however it does not affect the observed rate of LGRBs with current instruments (i.e. the number of LGRBs in the various mock samples presented in Sect.~\ref{subsec:pop_mock_samples}) since the majority of their bursts have luminosities above \Lmin.

We explored two scenarios for the intrinsic luminosity function: the first in which it is constant, and the second in which it is allowed to evolve with redshift.
In the second case we assume for simplicity that the slope of the luminosity function $p$ is constant and only \Lmin\ and \Lstar\ evolve with redshift as $(1+z)^{\kevol}$ as in \citet{Salvaterra2012}.
The redshift-evolving luminosity function is given by:
\begin{equation}\label{eq:lum_evol}
\phi(L,z) = \frac{1}{(1+z)^{\kevol}} \, \phi\left(\frac{L}{(1+z)^{\kevol}}, z=0\right) 
\end{equation}
where the $1/(1+z)^{\kevol}$ pre-factor comes from the condition that the probability density remain normalised to unity.
In this case, the values of the parameters quoted are always given for the de-evolved luminosity function, at $z = 0$.

\subsection{Intrinsic photon Spectrum}\label{subsec:pop_spectr}
There are a number of different functional forms used to fit GRB high energy spectra (typically 10 keV - 1 MeV) throughout the literature \citep[see e.g.][]{Preece2000,Kaneko2008,Goldstein2013,Gruber2014,Bhat2016}.
We decided to use the Band function \citep{Band1993} which provides the best fits for GRBs with high S/N in the GBM catalogue (see left pannels of Fig.~\ref{fig:Band_quality}) and whose shape has only 3 parameters: the low-energy slope $\alpha$, the high-energy slope $\beta$, and the peak energy \Ep.
The Band function is presented in more detail in App.~\ref{app:band}.

\subsubsection{Intrinsic peak energy distribution}
In the past, various authors have found relations between the spectral and energetic properties of GRBs and in particular between the peak energy \Ep\ of the GRB spectrum and the isotropic-equivalent energy \Eiso\ \citep{Amati2002,Amati2006,Lu2012} or luminosity \Liso\ \citep{Yonetoku2004,Yonetoku2010,Frontera2012}.
Some authors have suggested that these correlations are caused by strong selection effects \citep{Nakar2005,Band2005,Butler2007b,Shahmoradi2011,Heussaff2013}, while others have argued that selection effects do not suffice to explain the observed correlation \citep{Ghirlanda2008,Nava2008,Ghirlanda2012}.
For our models we tested two different scenarios regarding the \Ep\ distribution of LGRBs: a scenario with an intrinsic correlation between \Ep\ and \Liso\ and a scenario where the intrinsic \Ep\ distribution follows a log-Normal distribution independently of \Liso.

Several population models \citep[e.g.][]{Salvaterra2012,Pescalli2016,Lan2019}
assume that the observed correlation between the peak luminosity of a burst and its peak energy is intrinsic.
The correlation has 3 free parameters and can be written as:
\begin{equation}\label{eq:intr_correl}
\Ep = \Epo\,\left(\frac{L}{L_0}\right)^{\alphaA}
\end{equation}
where \Epo\ is drawn from a log-Normal distribution with scatter \sigmaEp, and $L_0$ is a constant fixed at $1.6 \times 10^{52}\,\ergs$.
The population models cited above use the values obtained by fitting observed samples of LGRBs and assumed to be valid for the intrinsic LGRB population; the most recent observed values from the extended BAT6 sample are $\Epo=309\pm6\,\keV$, $\alphaA=0.54\pm0.05$, $\sigmaEp=0.28$ \citep{Pescalli2016}.
In the intrinsic \EpL\ correlation scenario studied in the present paper, we
decided to relax this condition and leave the parameters of this relation free to vary to allow for an intrinsic \EpL\ correlation to be affected by additional selection effects to shape the observed correlation.

The independent log-Normal \Ep\ scenario assumes \Ep\ is independent of $L$ and follows a log-Normal distribution with 2 free parameters: the mean \Epo\ [keV] and the spread \sigmaEp\ [dex].
Note that the correlated scenario becomes the independent log-Normal scenario when \alphaA\ goes to 0.
The motivation behind this scenario is to see whether we can obtain an observed \EpL\ correlation without imposing any intrinsic correlation.

\subsubsection{Intrinsic spectral slopes distribution}
The spectral slopes $\alpha$ and $\beta$ are drawn from the catalogue of \citet{Gruber2014,Bhat2016}.
In particular, we chose the sample fulfilling their "good" criteria (i.e. with small errors, \citealt{Gruber2014}) adding the conditions that $2 < \beta$ and $\alpha < 2$.
In this way we have a diversity of realistic values for these parameters as opposed to fixing them as is commonly done in GRB population models \citep[e.g.][]{Wanderman2010,Salvaterra2012}.

\begin{figure}[!t]
\centering
\includegraphics[width=\linewidth]{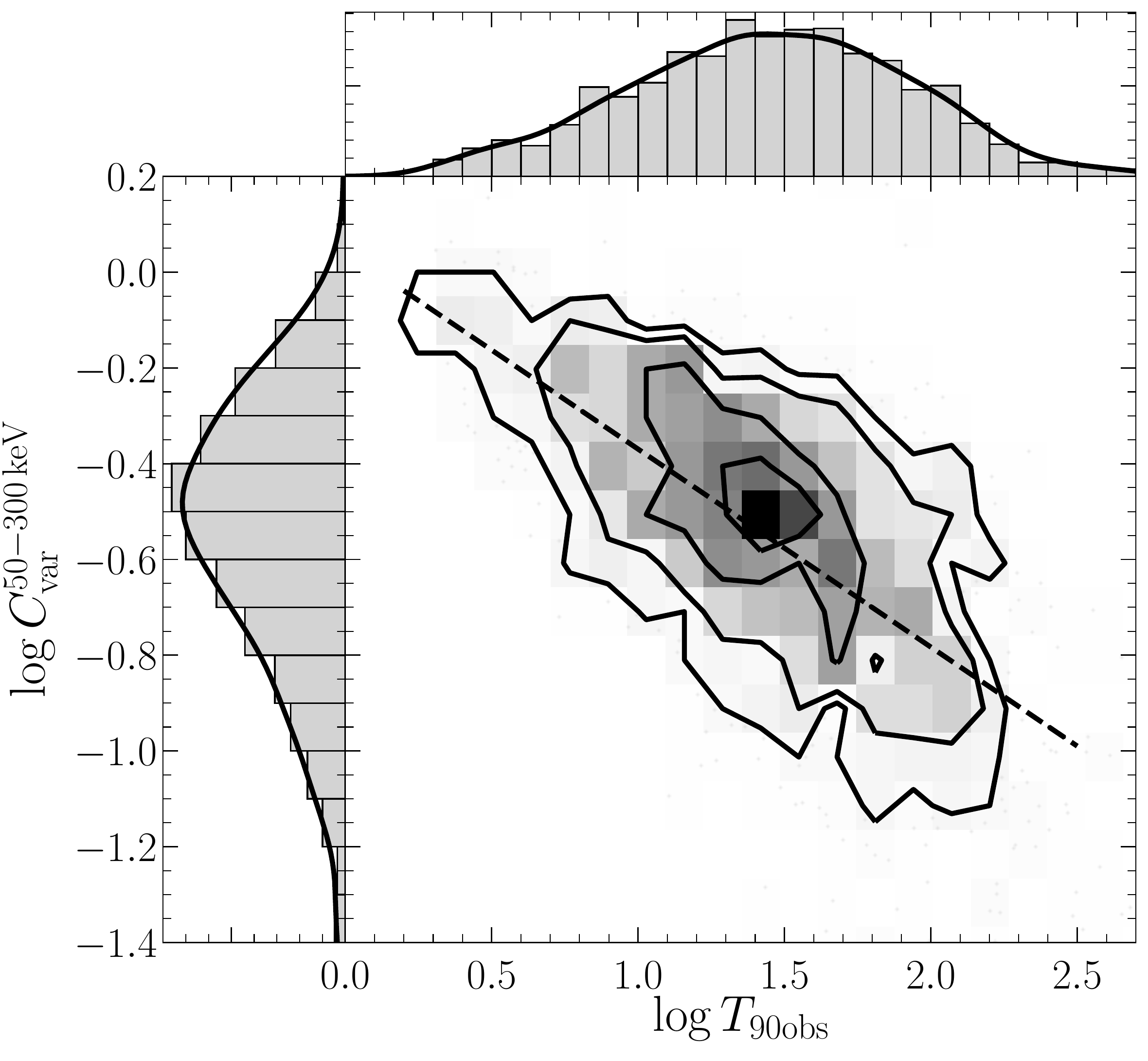}
\caption{Distribution of \Cvar\ and \Tnt\obs\ for the GBM catalogue of \cite{Bhat2016} for burst with $\Npksub{50}{300}{keV}\,\geq\,0.9\,\phscm$.
The data are binned in a 2D histogram whose shading represents the density of points; contours are also shown to guide the eye.
The black dashed line corresponds to the linear fit of Eq.~\ref{eq:Cvar_T90_fit}.
The black filled lines in the side histograms are the 1D Gaussian kernel density estimations of the data.}
\label{fig:Cvar}
\end{figure}

\subsection{Intrinsic duration and variability coefficient distributions}\label{subsec:pop_Cvar}
The intrinsic duration and variability coefficients are used solely for cross-checks and are drawn after the best fit parameters for a given scenario have already been found by MCMC (see Sect.~\ref{sec:results}).
This means that all mock samples with a selection based only on the peak flux have already been built, including the \bGBM\ sample of bright GBM bursts (cut at 0.9\,\phscm\ to avoid faint flux incompleteness, see Sect.~\ref{subsec:EpGBM}).
This allows us to compute the probability density function of the intrinsic duration \Tnt\ by first fitting the observed distribution of $\Tnt\obs$ from the GBM catalogue, cutting out LGRBs below 0.9\,\phscm, with a log-Normal distribution which yields a mean $\mu=1.45$ and standard deviation $\sigma=0.47$.
Then we assume the intrinsic distribution of \Tnt\ is log-Normal with the same spread and correct for cosmological time dilation as $\mu-\log(1+\bar{z}_{\rm GBM})$, where $\bar{z}_{\rm GBM}$ is the median redshift of our mock \bGBM\ sample.

The variability coefficient \Cvar\ is defined as the ratio of the mean luminosity $\bar{L}$ to the peak luminosity $L$.
It can be estimated from the GBM catalogue with:
\begin{equation}\label{eq:Cvar_fluence}
\Cvar=\frac{\bar{L}}{L}=\frac{\bar{N}}{\Npk}=\frac{\mathcal{N}}{\Tnt\obs\,\Npk}
\end{equation}
where $\mathcal{N}$ is the photon fluence (in units of [\phcm]) and \Npk\ is the peak photon flux (in units of [\phscm]) defined in Eq.~\ref{eq:pflx}.
The \Cvar-\Tnt\obs\ plane of the GBM catalogue with $\Npksub{50}{300}{keV} \geq 0.9~\phscm $ is shown in Fig.~\ref{fig:Cvar} in the case of a Band spectral fit using the flux and fluence values in the 50-300\,keV band.
We fit the correlation between \Cvar\ and \Tnt\obs\ with a linear regression and find:
\begin{equation}\label{eq:Cvar_T90_fit}
\log\,\Cvar=-0.413\,\log\Tnt\obs
\end{equation}
We crosschecked these values on different catalogues (BATSE 5B, \citealt{Goldstein2013}) and different energy bands (GBM 10-1000\,keV, \citealt{Bhat2016}) and find that the slope is within 10\% around $-0.4$.
We fit the decorrelated \Cvar\ with a log-Normal distribution yielding $\mu=0.04$ and $\sigma=0.22$.
Thus to obtain \Cvar\, we: (i) draw $z$ and \Tnt\ as mentioned above and calculate \Tnt\obs, (ii) draw log~$C_{\rm var}^{\rm decorr}$ from a log-Normal distribution with $\mu=0.04$ and $\sigma=0.22$, and (iii) calculate \Cvar\ from Eq.~\ref{eq:Cvar_T90_fit}.
Note that there can be some nonphysical values of \Cvar$>1$ which we set to 1 in our implementation.

\subsection{Mock samples}\label{subsec:pop_mock_samples}

\renewcommand{\arraystretch}{1.2}
\begin{table}
\caption{Summary of the various mock samples created for our population model.
The top part of the table corresponds to the samples used for fitting the observation constraints presented in Sect.~\ref{sec:obs_constraints}.
The bottom part of the table are additional samples used for predictions or additional cross-checks.
}
\label{tab:mock_samples}
\centering
\begin{tabular}{lcc}
\toprule
Sample name & \Eminobs\ - \Emaxobs\ & Threshold \\
\midrule
\BATSE & 50 - 300 keV  & 0.067~\phscm \\
\bGBM & 50 - 300 keV  & 0.9~\phscm \\
\eBATsix\ & 15 - 150 keV  & 2.6~\phscm \\
\cmidrule(lr){1-3}
\multirow{2}{*}{\textit{HETE2}} & 2 - 10 keV & 1~\phscm \\
&  30 - 400 keV & 1~\phscm \\

SHOALS & 15 - 150 keV & $1\times10^{-6}$~\ergcm\\
\bottomrule
\end{tabular}
\end{table}
\renewcommand{\arraystretch}{1.}

From the distributions presented above, and for a given set of parameters in the case of the adjusted distributions, we can randomly draw and calculate the properties and quantities given in Tab.~\ref{tab:pop_quant} for a large number of LGRBs \NGRB.
From this intrinsic population we can create several mock samples by applying different peak flux, or fluence cuts; these are summarised in Tab.~\ref{tab:mock_samples}.
The first three mock samples of Tab.~\ref{tab:mock_samples} are only based on a peak flux cut and therefore do not depend on \Tnt\ and \Cvar.
They are the ones used for comparison with the observational constraints presented in Sect.~\ref{sec:obs_constraints} in order to constrain the parameters of the population's distributions.
The SHOALS sample relies on a fluence criteria and thus also requires \Tnt\ and \Cvar; it is used as a cross-check (see Sect.~\ref{subsubsec:cross_check_SHOALS}) rather than a constraint for this reason.

\section{Observational Constraints}\label{sec:obs_constraints}

\begin{figure*}
\begin{center}
\includegraphics[width=\textwidth]{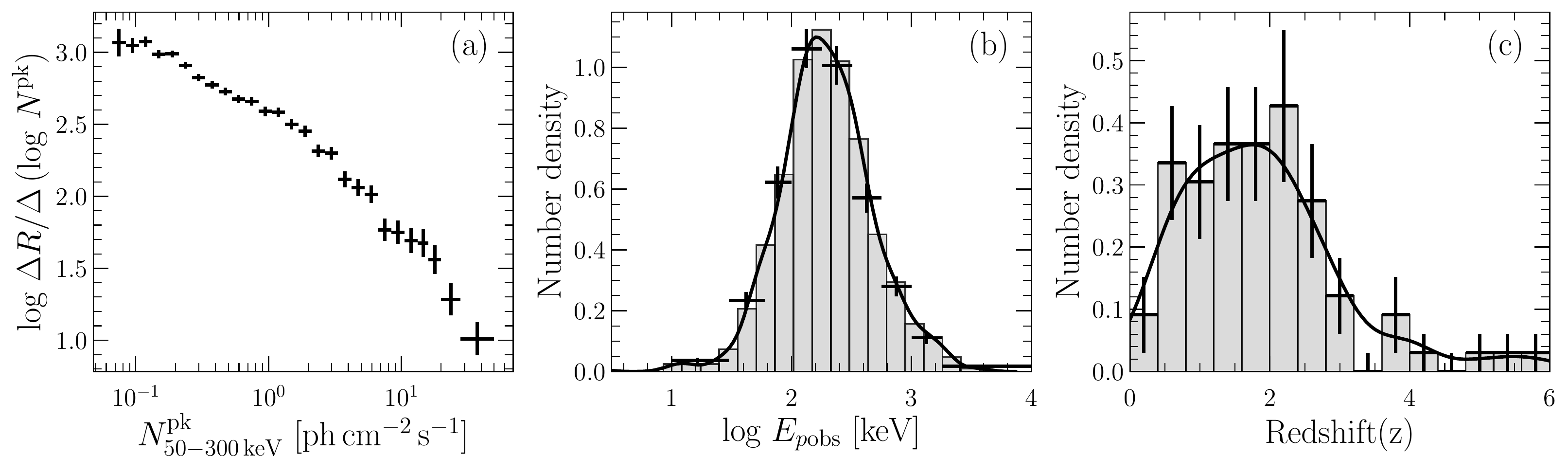}
\caption{(a) {\it Intensity constraint}: $\log{N}$-$\log{P}$ diagram built from the offline search of \BATSE\ LGRBs of \citet{Stern2001}, corrected for efficiency of detection at low fluxes.
(b) {\it Spectral constraint}: \Ep\ number density distribution from the GBM spectral catalogue \citep{Gruber2014} for long GRBs with \Npksub{50}{300}{keV} $> 0.9\,\phscm$. The actual histogram used in the fitting procedure is shown as black crosses while a finer-grained histogram is shown in light-grey to guide the eye. The black line is a Gaussian kernel density estimation of the data.
(c) {\it Redshift constraint}: Redshift number density distribution of the \eBATsix\ sample. The black line is a Gaussian kernel density estimation of the data.}
\label{fig:obs_constr}
\end{center}
\end{figure*}

In order to constrain our LGRB population model, we used carefully selected observational constraints covering all key aspects of the observed LGRB population. 
We used different missions and instruments to achieve this, taking advantage of the wide variety of observational data publicly available as of October 2018.
This use of samples from three different instruments is one new aspect of our population model with respect to others and requires that special attention be paid to the selection process.
Additionally, this implicitly assumes that the LGRBs from these various missions all come from the same underlying LGRB population, an assumption which is validated given the good quality fits described in Sect.~\ref{sec:results}.

\subsection{Intensity constraint}\label{subsec:Stern}
One of the most important constraints, in particular for the luminosity function of LGRBs, is one based on the intensity of the bursts.
In this instance, the intensity of the bursts can be assimilated to the peak flux, thus it becomes of interest to constrain the number of bursts at each peak flux, traditionally known as the $\log{N}$-$\log{P}$ diagram\footnote{To be rigorous, what we present in panel (a) of Fig.~\ref{fig:obs_constr} is a $\log{R}$-$\log{N}$ diagram, where $N$ is the peak photon flux \Npk\ and $R$ is the rate of GRBs per  $\Delta\log{\Npk}$\,bin; for simplicity we will refer to this as a $\log{N}$-$\log{P}$.}, represented in panel (a) of Fig.~\ref{fig:obs_constr}.

This diagram is a good way to estimate the peak isotropic-equivalent luminosity function of GRBs, however there is a difficulty residing in the fact that while peak fluxes are proportional to the luminosity, they also depend on redshift.
This means a burst with a high peak flux could be low luminosity at low redshift, or high luminosity at high redshift.
Fruitful studies \citep{Kommers2000,Stern2001} have focused on the turnover at low peak flux, trying to determine if it is real (i.e. due to a minimum luminosity of GRBs) or if it is caused by the lower efficiency of detectors at these fluxes.
In the case of \citet{Stern2001}, who went down to the lowest peak flux limit by conducting an off-line search of all \BATSE\ records, they conclude that there is no turnover down to a peak flux of about $\sim 0.1\,\phscm$.
Using the catalogue\footnote{\url{https://heasarc.gsfc.nasa.gov/W3Browse/gamma-ray-bursts/sterngrb.html}} from \citet{Stern2001}, we reconstructed a modified version of their original $\log{N}$-$\log{P}$ diagram, using wider bins towards high peak fluxes to insure at least 10 objects in each bin for proper Gaussian statistics to be applicable.

\subsection{Spectrum constraint}\label{subsec:EpGBM}
\begin{figure}
\begin{center}
\includegraphics[width=\linewidth]{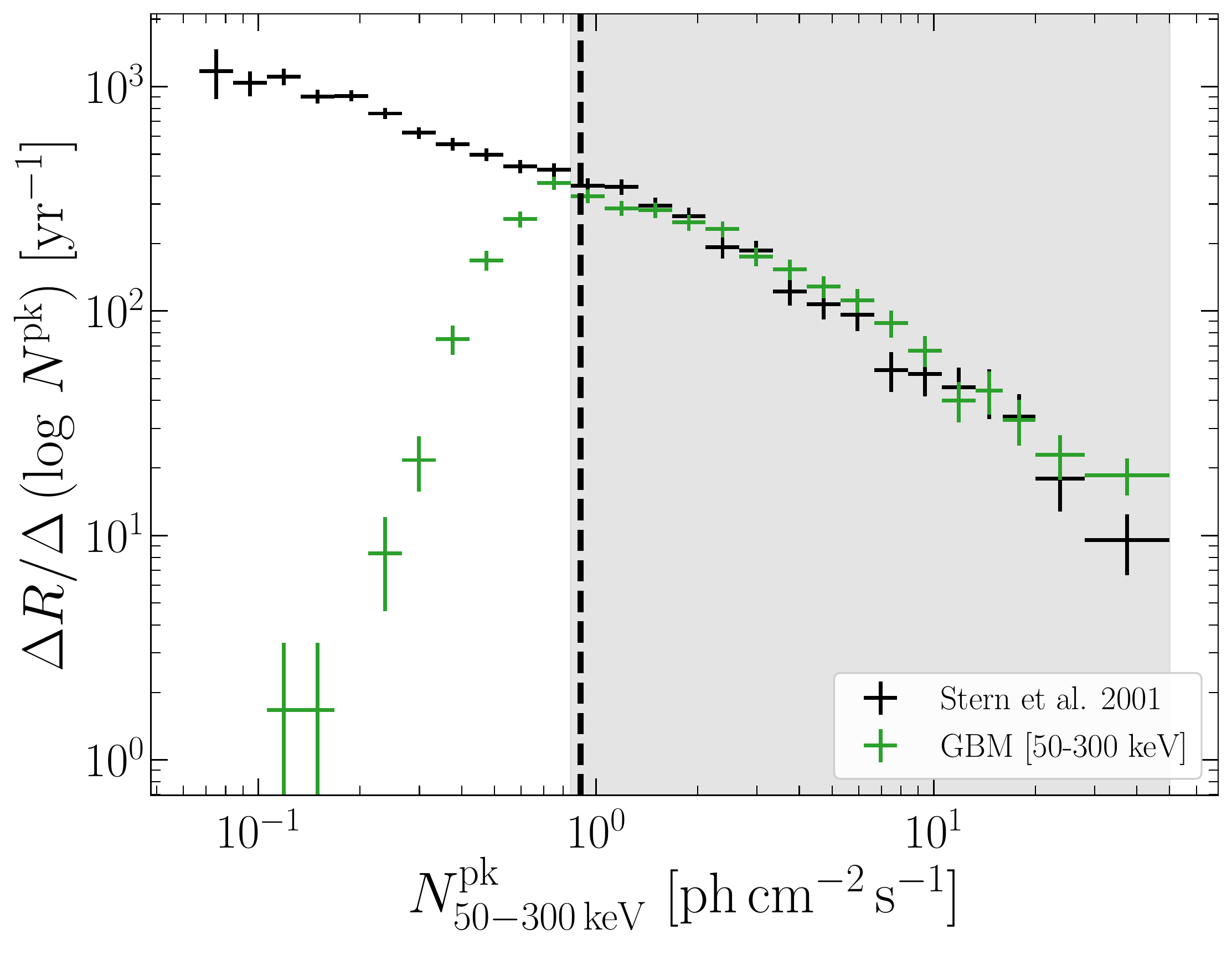}
\end{center}
\caption{The efficiency-corrected $\log{N}$-$\log{P}$ diagram for \BATSE\ from \citet{Stern2001} in black.
The $\log{N}$-$\log{P}$ of GBM is shown in green, adjusted to the one of \BATSE\ by multiplying by a constant whose value is obtained by minimizing the \Chisq\ between the two histograms over the grey shaded region.
The black dashed vertical line represents the \Npk\ cut to avoid biases due to faint flux incompleteness for our spectral constraint.}
\label{fig:EpGBM_selection}
\end{figure}

In order to constrain the spectral properties of our intrinsic population we focused on a quantity that is fundamental in defining the GRB prompt spectrum: the peak energy \Ep.
Similarly to the $\log{N}$-$\log{P}$ distribution, the \Ep\obs\ distribution is the result of the intrinsic \Ep\ distribution, convolved with redshift which raises the same aforementioned problems.
In addition to this, there is the issue of properly measuring \Ep, which is difficult for instruments with a narrow energy band (e.g. \textit{Swift}/BAT 15-150\,keV).
We decided to use data from the 3rd \textit{Fermi}/GBM catalogue\footnote{\url{https://heasarc.gsfc.nasa.gov/W3Browse/fermi/fermigbrst.html}} \citep{Gruber2014,vonKienlin2014,Bhat2016} since GBM is an instrument with a large sample ($\geq$1500 GRBs) and a broad spectral bandwidth (10~keV - 30~MeV).
In order to have a clean comparison sample we created the $\log{N}$-$\log{P}$ diagram for the GBM sample and compared it to the one from \citet{Stern2001}, shown in Fig.~\ref{fig:EpGBM_selection}, using the GBM peak flux in the 50-300\,keV band.
Since a thorough study on the efficiency of the GBM detectors is not yet available, we normalized the $\log{N}$-$\log{P}$ from GBM to the corrected one from \citet{Stern2001} by multiplying the GBM histogram by a constant and performing \Chisq\ minimization on the bright bins (shown within the grey shaded area in Fig.~\ref{fig:EpGBM_selection}) to find the best value.
This normalization yields an average exposure factor of $f_\mathrm{exp}\simeq 0.57$ for the GBM sample (fraction of observing time multiplied by the fraction of sky observed, see Eq.~\ref{eq:av_exp}) slightly higher than the value for \BATSE\ of 0.46 \citep{Salvaterra2012,Goldstein2013}.
We then cut the GBM sample at 0.9 \phscm, indicated by the black dashed vertical line in Fig.~\ref{fig:EpGBM_selection}, below which the $\log{N}$-$\log{P}$ diagrams of GBM and \citet{Stern2001} start to diverge.
By doing this procedure we are ensuring an \Ep\obs\ distribution that is unbiased from faint flux incompleteness, at the price of sample size.
The resulting \Ep\obs\ distribution is shown in panel (b) of Fig.~\ref{fig:obs_constr}.

\subsection{Redshift constraint}\label{subsec:eBAT6}
As illustrated by the previous sections, the distance of GRBs is inherently intertwined with any observable and unfortunately the majority of GRBs do not have a measured redshift.
To date, there are over 500\footnote{\url{http://www.mpe.mpg.de/~jcg/grbgen.html}} GRBs with measured redshift, however it is not possible to simply use all GRBs with a redshift since redshift distributions are often plagued with strong selection effect and biases which are difficult to model \citep[see however][]{Wanderman2010}.
For instance, the ability to measure a redshift for GRBs relies fundamentally on the capacity to locate it, which biases this distribution against so-called \textit{dark} bursts \citep{Greiner2011,Melandri2012} which exhibit highly extinguished optical afterglows.
Another selection effect, called the redshift desert, is due to the fact that most emission and common absorption lines are shifted outside the window of optical spectrographs around $z\sim 2$, although the advent of newer spectrographs such as X-Shooter \citep{Vernet2011} have mostly remedied this.
It is thus crucial to use a well-controlled redshift distribution to avoid biasing the intrinsic LGRB population, even at the cost of sample size; we therefore used the redshift distribution from the extended BAT6 sample.

The \textit{Swift}/BAT6 sample \citep{Salvaterra2012} is a complete sample of \textit{Swift} LGRBs with a selection based on the peak flux $\Npksub{15}{150}{\keV} > 2.6$ \phscm\ and favorable observing conditions \citep{Jakobsson2006} .
These conditions are chosen so that they increase the chance of redshift recovery without biasing the redshift distribution.
They are based on criteria which do not depend on the redshift of the LGRB:
\begin{itemize}
    \renewcommand\labelitemi{--}
    \item The burst must be well localized by \textit{Swift}/XRT and the information was distributed quickly
    \item There is low galactic foreground extinction (\Av\ < 0.5)
    \item The burst declination is between -70\degree\ and +70\degree
    \item The burst's angular distance to the sun is greater that 55\degree
    \item There are no nearby bright stars
\end{itemize}

This results in 58 LGRBs for the original BAT6, later extended to 99 LGRBs by \citet{Pescalli2016}.
This extended BAT6 sample (\eBATsix) has 82 bursts with redshifts, yielding an 83\% redshift completeness, and its redshift distribution is shown in panel (c) of Fig.~\ref{fig:obs_constr}.
It should be noted that while the eBAT6 bursts are statistically representative of the population of {\it Swift} bursts with $\Npksub{15}{150}{\keV} > 2.6$ \phscm, this population constitutes only the brightest 25\% of {\it Swift} bursts.

\subsection{Event rates for our mock samples}\label{subsec:pop_norm}

One of the advantages of using the \BATSE\ $\log{N}$-$\log{P}$ of \citet{Stern2001} is that they derived a detection efficiency for all bursts within the field of view.
Applying this efficiency correction while also taking into account the mean observed solid angle of the sky $\mOmega{BATSE} = 8.17~\rm{sr}$ and the live time of the search $\Tlive{BATSE} = 6.54~\rm{yr}$ \citep{Goldstein2013}, we can use this corrected $\log{N}$-$\log{P}$ to obtain the total observed all-sky LGRB rate:
$$
\Rate{tot}{BATSE}=1170\,(\pm\,40)~\rm LGRB\,\yr\,in\,4\,\pi
$$
for BATSE LGRBs (i.e. brighter than $0.067~\phscm$ in the 50-300\,keV range).
We can use \Rate{tot}{BATSE} to normalise our population with:
$$
\Tsim = \frac{\Nb{BATSE}}{\Rate{tot}{BATSE}}~~~~~~~~~~[\rm yr]
$$
where \Nb{BATSE} is the number of LGRBs in our mock \BATSE\ sample and \Tsim\ represents the time it would take to generate \Nb{BATSE} LGRBs with an LGRB rate of \Rate{tot}{BATSE}.
Defining the intrinsic rate of all LGRBs above \Lmin\ as:
$$
\Rate{}{intr} = \frac{\NGRB}{\Tsim}~~~~~~~~~~~[\yr]
$$
where \NGRB\ is the number of LGRBs in our simulated population, we can obtain \nGRBo\ with:
\begin{equation}
\Rate{}{intr} = \nGRBo\,\int_0^{z_{\rm max}}\frac{f(z;a,b,z_m)}{1+z} \frac{\dd V}{\dd z} \dd z
\end{equation}
where $z_{\rm max}$ is a maximum redshift, set to 20 in our implementation.
We can also calculate the total rate of LGRBs predicted by our population for a given sample:
$$
\Rate{tot}{sample} = \frac{\Nb{sample}}{\NGRB}\,\Rate{}{intr}\, ~~~~~~~[\yr]
$$
The observed rate of LGRBs for a given sample is then simply the total rate times the average exposure factor $f_{\rm exp}$:
\begin{equation}\label{eq:av_exp}
\begin{split}
\Rate{obs}{sample} & = f_{\rm exp}\,\Rate{tot}{sample}~~~~~~~~~~~~~~~~~~~~~~~~~~[\yr]  \\
& = \frac{\langle  \Omega \rangle}{4\,\pi}\,\frac{T_{\rm live}}{T_{\rm total}}\,\Rate{tot}{sample}  
\end{split}
\end{equation}
where $\langle  \Omega \rangle$ is the average angle of the sky observed by the instrument and $T_{\rm live}$ and $T_{\rm total}$ are the live time and total time of observation of the instrument respectively.

\subsection{Additional cross-checks}\label{subsec:cross_check}
In addition to the aforementioned constraints used for fitting, we control that the distributions of the spectral slopes $\alpha$ and $\beta$ of the GBM sample generated by our model are consistent with the observed distributions and include two other observables to help discriminate scenarios with similar likelihood.
A brief description is given below but see Sect.~\ref{subsec:disc_EpL_correl} and \ref{subsec:disc_cosmic_evolution} for more details.

\subsubsection{The \textit{Swift}/eBAT6 Ep-L plane}

\begin{figure}[ht!]
\begin{center}
\includegraphics[width=\linewidth]{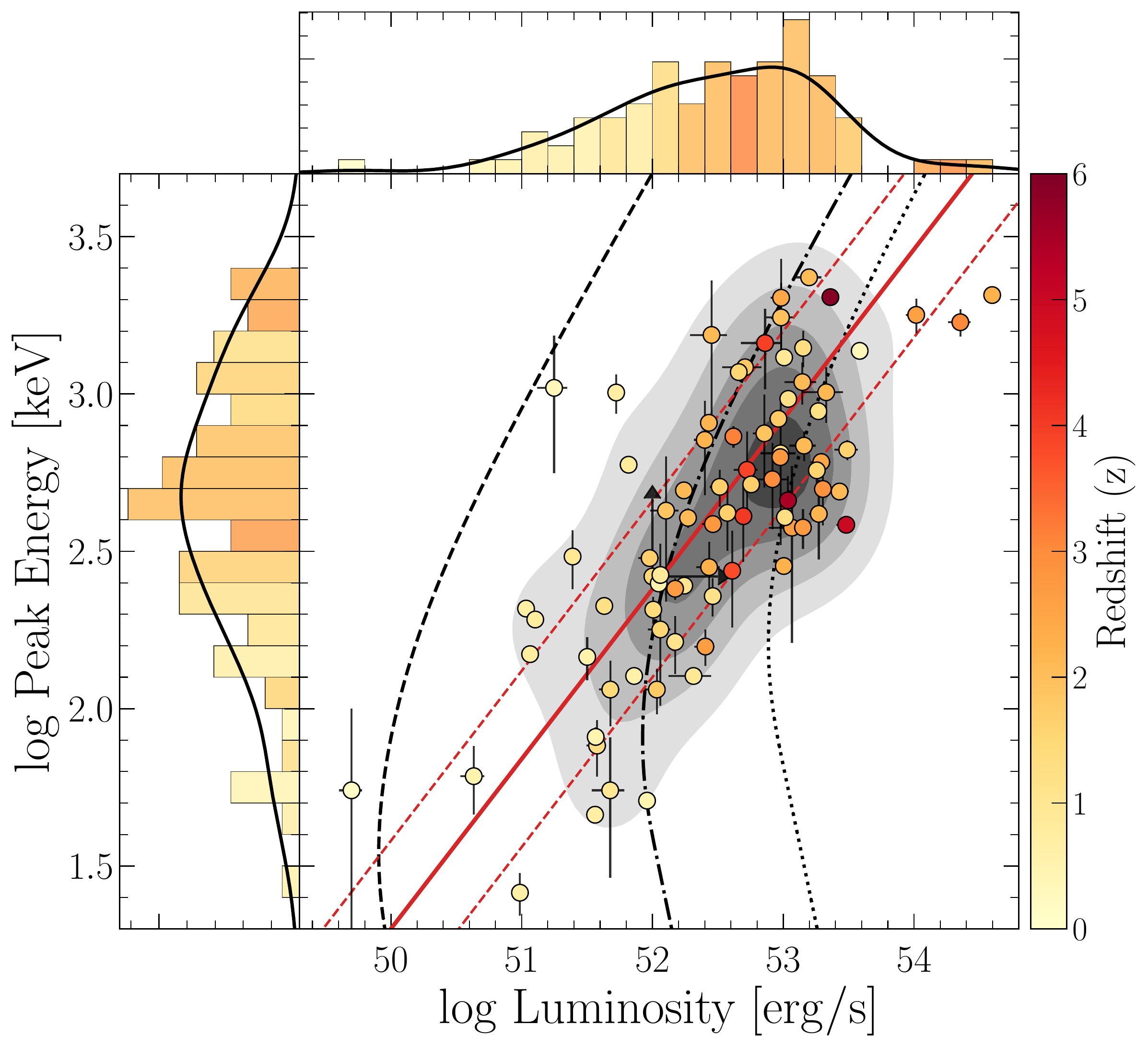}
\caption{
\EpL\ plane for the \eBATsix\ sample.
The individual data points are colour-coded by redshift; the filled red line represents the \Ep-\Liso\ relation found by \citet{Pescalli2016} for this sample and the dashed red lines represents the scatter.
A 2D Gaussian kernel density estimation of the data is shown as grey shaded contours.
The black dashed, dot-dashed and dotted lines represent the detection threshold for $\Npksub{15}{150}{\keV} = 2.6$ \phscm\ and a fixed Band spectrum ($\alpha=0.6$, $\beta=2.5$) at redshifts 0.3, 2, and 5 respectively.
The side histograms represent the binned data and the colour of each bin represents the median redshift in that bin following the colour-code of the central panel; the black curve is the 1D Gaussian kernel density estimation.
}
\label{fig:eBAT6_EpL}
\end{center}
\end{figure}

The \eBATsix\ sample is used as a redshift constraint, but there is more information to be extracted from this complete sample.
More specifically, we use the \EpL\ plane since it contains information about the observed correlation between the isotropic-equivalent luminosity of the bursts and their peak energy.
This is relevant, in particular when trying to distinguish between scenarios with intrinsic correlated or independent log-Normal \Ep\ distributions (see Sect.~\ref{subsec:EpGBM}).
Figure~\ref{fig:eBAT6_EpL} shows the \EpL\ plane for the \eBATsix\ sample (data is from \citealt{Pescalli2016}).
Fitting the \EpL\ plane of the extended BAT6 sample, \cite{Pescalli2016} derived $\Epo=309\pm6\,\keV$, $\alphaA=0.54\pm0.05$, $\sigmaEp=0.28$, their fit is represented by the blue line in Fig.~\ref{fig:eBAT6_EpL}.
It should be noted that $\sim25\%$ of the original BAT6 sample have peak energies only determined by \textit{Swift}/BAT, which can be inaccurate for determining spectral parameters due to its small bandwidth (15-150$~\keV$).
This could potentially impact the \Ep\ distribution of the sample, although a precise quantification of this effect has not yet been determined \citep[but see][]{Nava2012}.

\subsubsection{The SHOALS redshift distribution}\label{subsubsec:cross_check_SHOALS}
The \textit{Swift} Gamma-Ray Burst Host Galaxy Legacy Survey (SHOALS, \citealt{Perley2016b,Perley2016}) is the largest unbiased sample of LGRB host galaxies to date with 119 objects and a 92\% redshift completeness.
The selection is based on a fluence cut, a quantity which is not straightforwardly calculated by our model since we do not simulate light curves for our LGRBs.
It can however be calculated from the two additional properties $\Tnt$\ and $\Cvar$ as shown in Eq.~\ref{eq:flnc}.
Because of the additional assumptions needed to create the distributions of the duration and the variability indicator, we decided to use only \eBATsix\ as our primary redshift constraint and to perform a cross-check a posteriori to make sure the best fit populations also adequately represent SHOALS.

\section{Results}\label{sec:results}

\renewcommand{\arraystretch}{1.4}
\begin{table*}
\caption{Best fit parameter values in the case of a log-Normal \Ep\ scenario. Parameter without errors were fixed during the exploration.}
\label{tab:pop_result_LN}
\centering
\begin{adjustbox}{max width=\textwidth}
\begin{tabular}{lccccccccc}
\toprule
\multirow{2}{*}{Name} & \nGRBo  & \multicolumn{3}{c}{Luminosity Function} & \multicolumn{2}{c}{Peak Energy Distribution} & \multicolumn{3}{c}{Redshift Distribution}\\ \cmidrule(lr){3-5} \cmidrule(lr){6-7} \cmidrule(lr){8-10}
 &[\yrGpc] & log \Lstar & $p$ & \kevol & log \Epo & \sigmaEp & $z_m$ & $a$ & $b$ \\ \toprule
$\nGRB \propto$ CSFRD & $1.00^{+0.01}_{-0.01}$ & $52.4^{+0.1}_{-0.1}$ & $1.45^{+0.02}_{-0.02}$ &  $1.6^{+0.1}_{-0.1}$ & $2.86^{+0.02}_{-0.02}$ & $0.47^{+0.02}_{-0.02}$ & 1.9                    &  1.1                   &  $-0.57$                         \\
No Evolution   & $1.08^{+0.10}_{-0.10}$ & $53.0^{+0.1}_{-0.1}$ & $1.36^{+0.04}_{-0.06}$ &      0.0                & $2.84^{+0.02}_{-0.02}$ &  0.45                  & $2.2^{+0.1}_{-0.1}$ & $1.2^{+0.1}_{-0.1}$ & $-0.17^{+0.13\dagger}_{-0.06}$ \\
Mild Evolution  & $1.11^{+0.05}_{-0.06}$ & $52.7^{+0.1}_{-0.1}$ & $1.34^{+0.03}_{-0.03}$ &      0.5                & $2.84^{+0.02}_{-0.02}$ &  0.45                  & $2.1^{+0.1}_{-0.1}$ & $1.0^{+0.1}_{-0.1}$ & $-0.19^{+0.03}_{-0.05}       $ \\
Evolution    & $1.30^{+0.06}_{-0.05}$ & $52.6^{+0.1}_{-0.1}$ & $1.41^{+0.02}_{-0.02}$ &      1.0                & $2.84^{+0.02}_{-0.02}$ &  0.45                  & $2.2^{+0.1}_{-0.1}$ & $0.9^{+0.1}_{-0.1}$ & $-0.52^{+0.16\dagger}_{-0.06}$ \\
Strong Evolution   & $1.48^{+0.05}_{-0.07}$ & $52.2^{+0.1}_{-0.1}$ & $1.50^{+0.02}_{-0.02}$ &      2.0                & $2.84^{+0.02}_{-0.02}$ &  0.45                  & $2.1^{+0.1}_{-0.1}$ & $0.7^{+0.1}_{-0.1}$ & $-0.62^{+0.05}_{-0.08}       $ \\ \bottomrule                   
\end{tabular}
\end{adjustbox}
\begin{tablenotes}
\tiny{
    \item $\dagger$ The parameter $b$ in these two cases has a multi-peaked marginalized posterior distribution; the median and 1~$\sigma$ errors reported here are not necessarily representative of the best fitting value but are quoted for simplicity. 
}
\end{tablenotes}
\end{table*}

\begin{table*}
\centering
\caption{Same as Tab.~\ref{tab:pop_result_LN} but for the intrinsic \EpL\ correlation scenario.}
\label{tab:pop_result_A}
\begin{adjustbox}{max width=\textwidth}
\begin{tabular}{lcccccccccc}
\toprule
\multirow{2}{*}{Name}&
\nGRBo & \multicolumn{3}{c}{Luminosity Function} & \multicolumn{3}{c}{Peak Energy Distribution} & \multicolumn{3}{c}{Redshift Distribution} \\\cmidrule(lr){3-5} \cmidrule(lr){6-8} \cmidrule(lr){9-11}
 &[\yrGpc] & log \Lstar & $p$ & \kevol & log \Epo & \sigmaEp & \alphaA & $z_m$ & $a$ & $b$ \\ \toprule
 $\nGRB \propto$ CSFRD  &
 $0.69^{+0.01}_{-0.01}$ & $52.5^{+0.1}_{-0.1}$ & $1.53^{+0.03}_{-0.03}$ &  $2.1^{+0.1}_{-0.1}$ & $2.80^{+0.03}_{-0.03}$ & $0.44^{+0.02}_{-0.02}$ & $ 0.25^{+0.05}_{-0.05}$ &  1.9                   &  1.1                   &  $-0.57$                  \\
 No Evolution    &
 $0.77^{+0.05}_{-0.05}$ & $53.3^{+0.1}_{-0.2}$ & $1.44^{+0.03}_{-0.06}$ &         0.0             & $2.79^{+0.03}_{-0.03}$ & $0.43^{+0.02}_{-0.02}$ & $ 0.30^{+0.05}_{-0.05}$ & $2.2^{+0.1}_{-0.1}$ & $1.35^{+0.1}_{-0.1}$ & $-0.18^{+0.02}_{-0.02}$ \\
 Mild Evolution  & 
 $0.73^{+0.21}_{-0.06}$ & $53.0^{+0.2}_{-0.2}$ & $1.42^{+0.07}_{-0.07}$ &         0.5             & $2.81^{+0.05}_{-0.06}$ & $0.44^{+0.03}_{-0.06}$ & $ 0.25^{+0.15}_{-0.15}$ & $2.1^{+0.1}_{-0.1}$ & $1.25^{+0.15}_{-0.15}$ & $-0.20^{+0.04}_{-0.06}$ \\
 Evolution    &
 $0.72^{+0.05}_{-0.05}$ & $52.9^{+0.2}_{-0.2}$ & $1.47^{+0.05}_{-0.05}$ &         1.0             & $2.78^{+0.03}_{-0.05}$ & $0.42^{+0.03}_{-0.03}$ & $ 0.35^{+0.1}_{-0.1}$ & $2.1^{+0.1}_{-0.1}$ & $1.2^{+0.1}_{-0.1}$ & $-0.27^{+0.06}_{-0.04}$ \\
 Strong Evolution  &  
 $0.79^{+0.06}_{-0.06}$ & $52.5^{+0.1}_{-0.1}$ & $1.52^{+0.03}_{-0.03}$ &         2.0             & $2.80^{+0.04}_{-0.04}$ & $0.44^{+0.03}_{-0.03}$ & $ 0.25^{+0.1}_{-0.1}$ & $2.1^{+0.1}_{-0.1}$ & $0.9^{+0.1}_{-0.1}$ & $-0.60^{+0.07}_{-0.06}$ \\ \bottomrule
\end{tabular}
\end{adjustbox}
\end{table*}
\renewcommand{\arraystretch}{1.}

In order to find the best fit parameters, we used a Bayesian framework with an indirect likelihood and a Markov Chain Monte Carlo (MCMC) approach to explore the parameter space; the details are presented in App.~\ref{app:stats}.
Due to computational considerations we tried to reduce the number of free parameters as much as possible in a justifiable way.
For instance, in the independent log-Normal \Ep\ scenario, we noticed the value of the standard deviation \sigmaEp\ did not move significantly from 0.45, we therefore decided to fix it at this value.
Moreover, we found a strong degeneracy between the luminosity evolution parameter \kevol\ and the slopes of the redshift distribution; we thus fixed \kevol\ to 0, 0.5, 1.0 and 2.0 and left the redshift distribution free to vary.
In the following, these 4 cases will be referred to as "no evolution", "mild evolution", "evolution" and "strong evolution" of the luminosity function respectively.
The full set of parameters explored and the ranges used for their flat priors are shown in Tab.~\ref{tab:priors}.

\begin{figure*}
\begin{center}
\includegraphics[width=\textwidth]{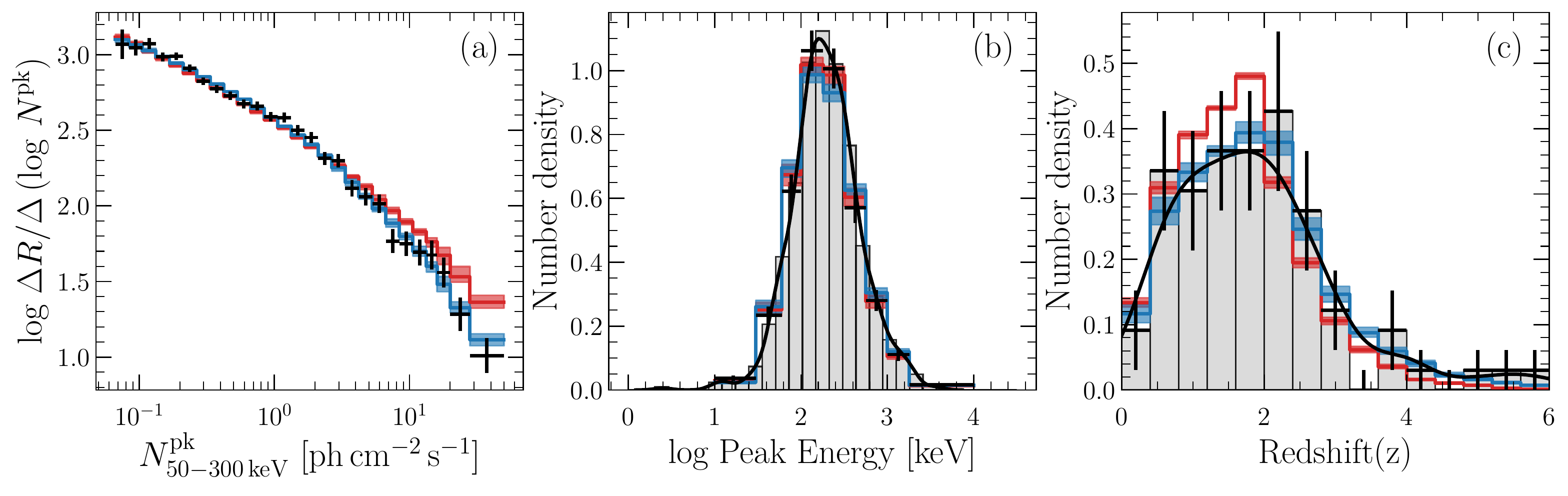}
\caption{Best fits to the observational constraints from models with intrinsic \EpL\ correlations and a luminosity function that does not evolve with redshift ($\kevol = 0$).
The red curve corresponds to the case where the LGRB rate follows the shape of cosmic star formation rate density and the blue curve corresponds to the case where the LGRB efficiency is free to evolve with redshift.
The black curves and grey histograms are the observational constraints described in Fig.~\ref{fig:obs_constr}.
This figure is discussed in Sect.~\ref{sec:results}.}
\label{fig:fp_A_nSFR}
\end{center}
\end{figure*}

The first result is that for a population with a non-evolving luminosity function ($\kevol=0$) and a constant LGRB efficiency (i.e. a redshift distribution that follows the shape of the cosmic star formation rate density), we cannot find a set of parameters that yield a good fit to the data.
Indeed, as shown in red in Fig.~\ref{fig:fp_A_nSFR} for the intrinsic \EpL\ correlation scenario, a population with no evolution of the luminosity function and a constant LGRB efficiency over-predicts the number of bright and low-redshift LGRBs.
This is an important result connected to the physics of LGRB progenitors and was already found by \citet{Daigne2006} after two years of \textit{Swift} observations.
It was then confirmed by following studies \citep{Wanderman2010,Salvaterra2012} and we confirm it again on larger, more precise samples and find that it is true both in the independent log-Normal \Ep\ scenario and the scenario with intrinsic \EpL\ correlations.
Our other models can then be used to better understand the underlying evolution (efficiency and/or luminosity) responsible for this discrepancy.

\begin{figure*}
\begin{subfigure}{.5\textwidth}
    \centering
    \includegraphics[width=\linewidth]{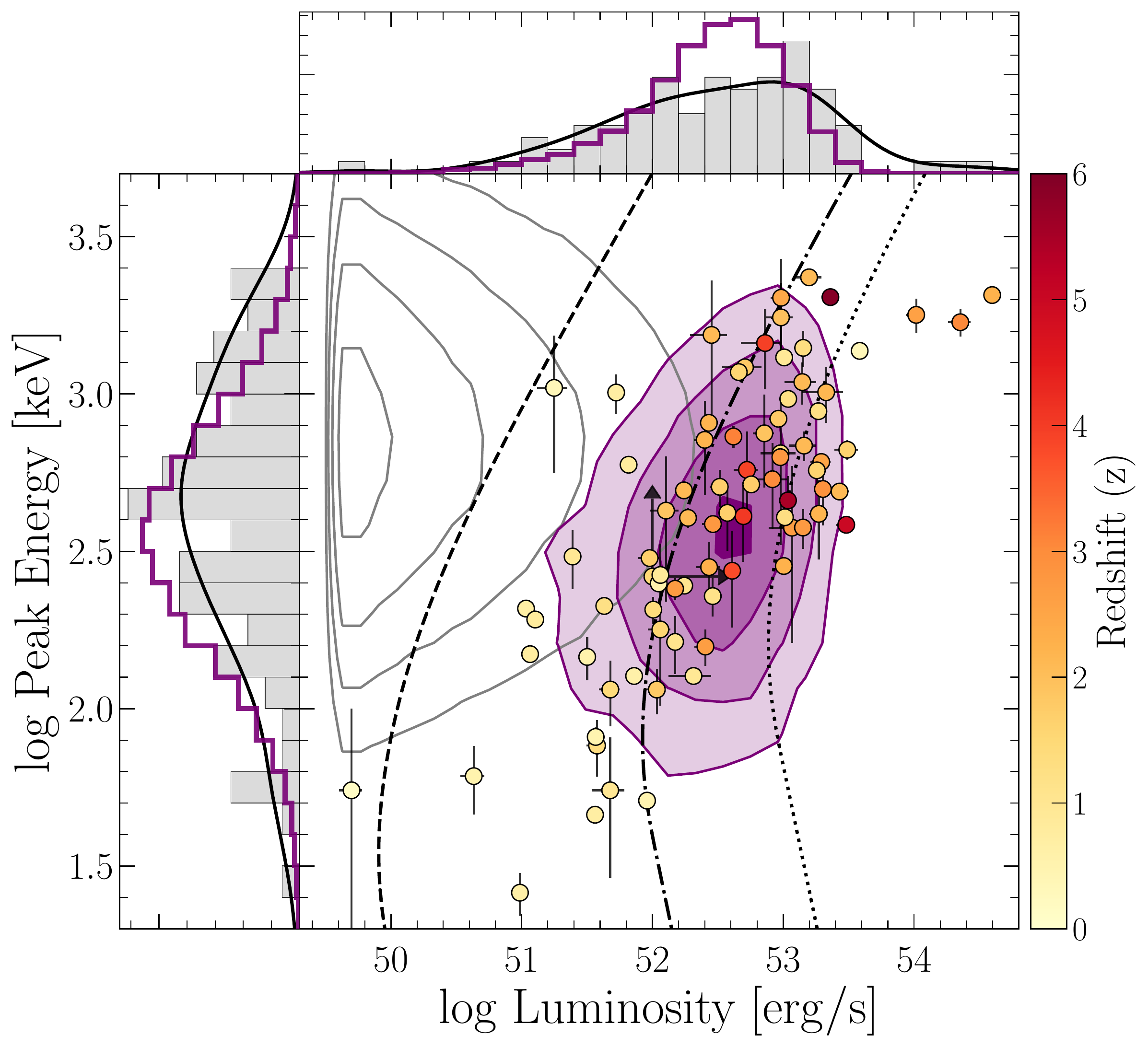}
    \caption{Log-Normal \Ep\ scenario with $\kevol = 0$}
    \label{fig:EpL_k0_LN_nF}
\end{subfigure}
\begin{subfigure}{.5\textwidth}
    \centering
    \includegraphics[width=\linewidth]{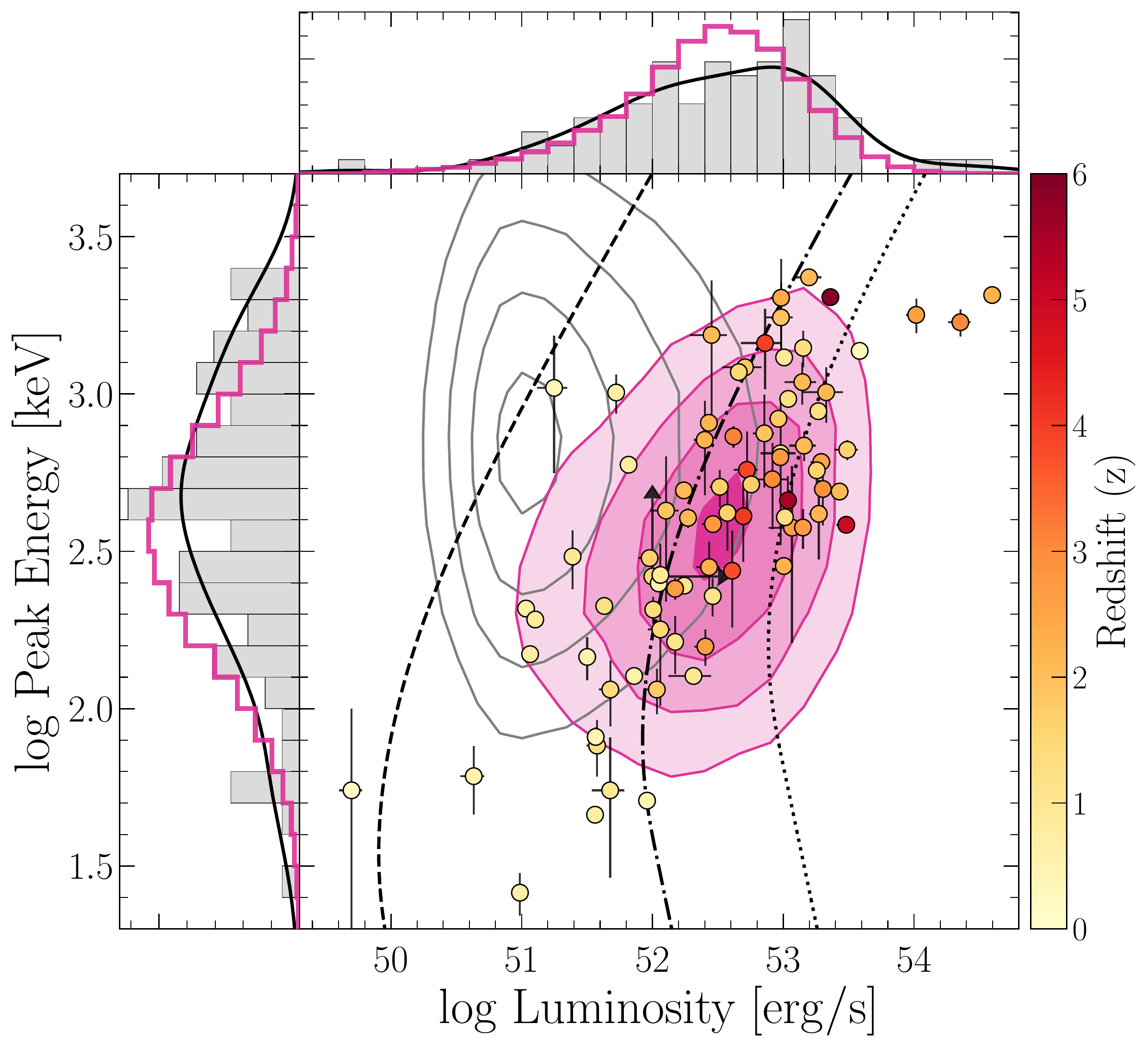}
    \caption{Log-Normal \Ep\ scenario with $\kevol = 2$}
    \label{fig:EpL_k2_LN_nF}
\end{subfigure}
\begin{subfigure}{.5\textwidth}
    \centering
    \includegraphics[width=\linewidth]{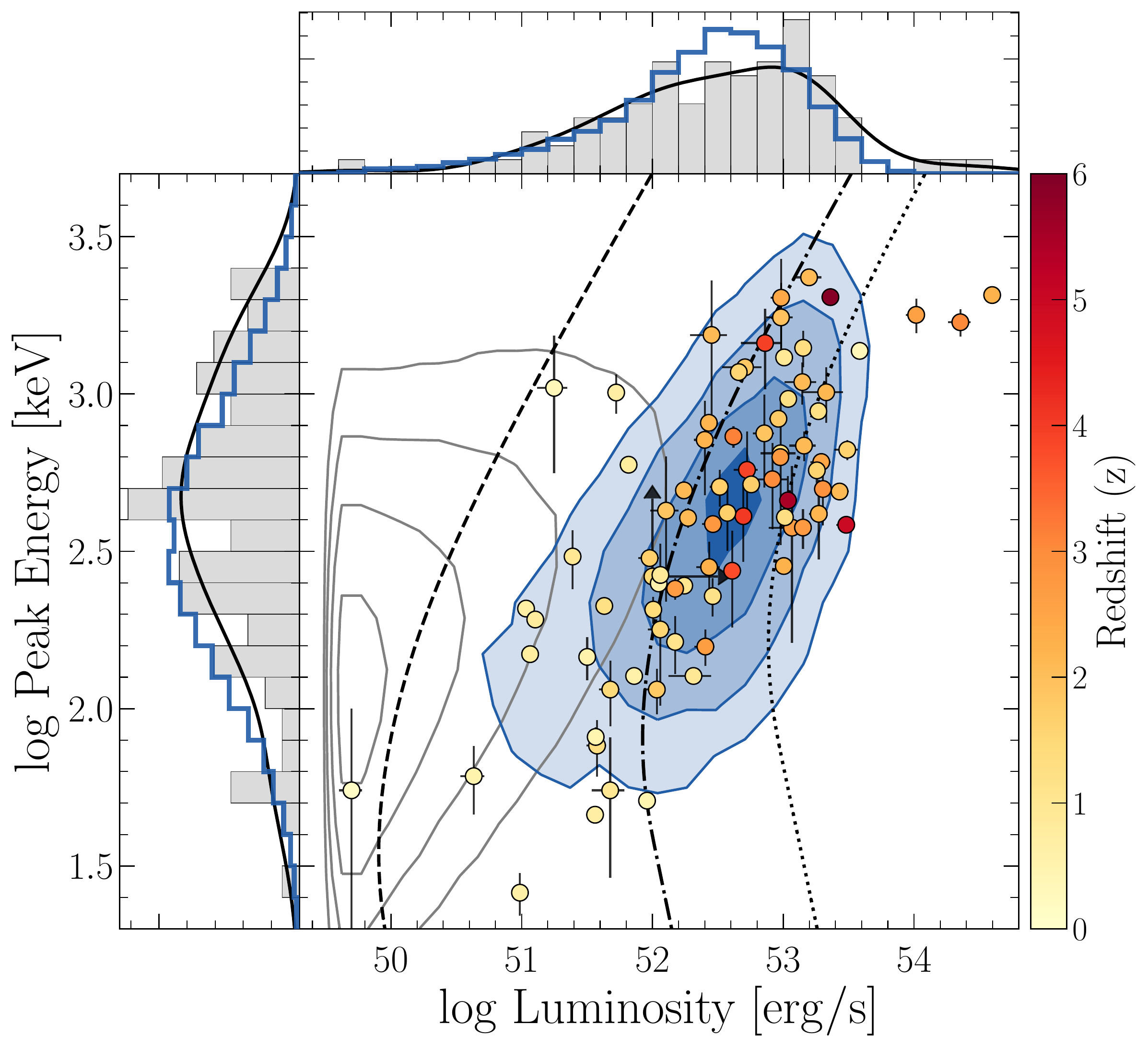}
    \caption{Intrinsic \EpL\ correlation scenario with $\kevol = 0$}
    \label{fig:EpL_k0_A_nF}
\end{subfigure}
\begin{subfigure}{.5\textwidth}
    \centering
    \includegraphics[width=\linewidth]{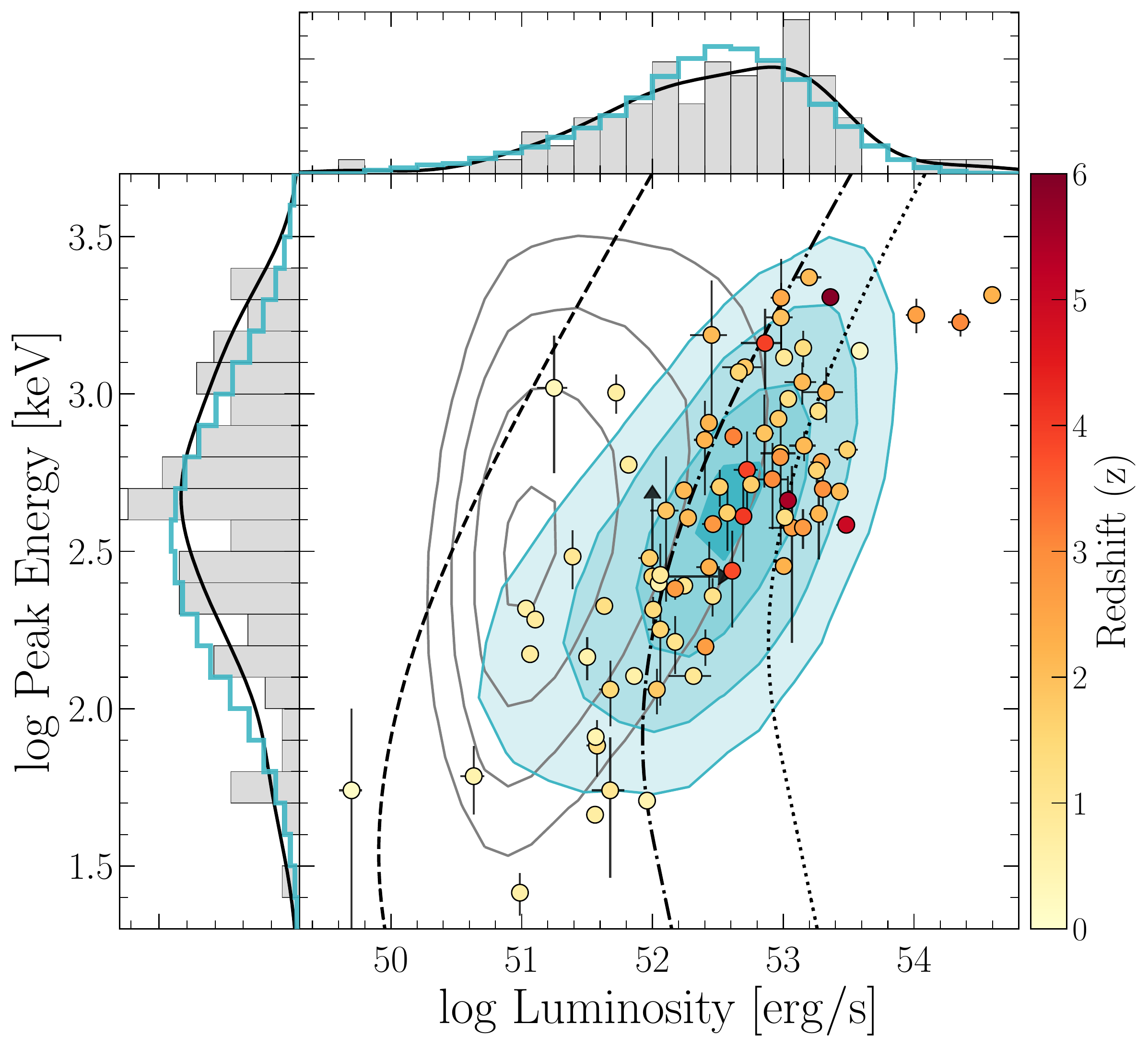}
    \caption{Intrinsic \EpL\ correlation scenario with $\kevol = 2$}
    \label{fig:EpL_k2_A_nF}
\end{subfigure}
\caption{The \eBATsix\ \EpL\ plane for models with no redshift evolution of the luminosity function (a, c) and with strong redshift evolution of the luminosity function (b, d).
The top panels (a, b) correspond to the independent log-Normal \Ep\ scenarios while the bottom panels (c, d) correspond to intrinsic \EpL\ correlation scenarios.
The predictions for the intrinsic LGRB population is shown as grey contours and the predictions for the mock eBAT6 sample are shown in coloured filled contours.
A description of the observed data points, the side histograms and the dashed, dot-dashed and dotted curves of the figure is presented in Fig~\ref{fig:eBAT6_EpL}.}
\label{fig:EpL_all}
\end{figure*}

Building in complexity and allowing either the redshift distribution parameters or \kevol\ to vary, good fits to the data are found; the best fit parameters for each case are reported in Tab.~\ref{tab:pop_result_LN} for the independent log-Normal \Ep\ scenario and in Tab.~\ref{tab:pop_result_A} for the intrinsic \EpL\ correlation scenario; the quality of the fits in each case is comparable.
\begin{itemize}
\item The scenarios where the redshift distribution is fixed proportional to the cosmic star formation rate density find strong evolution of the luminosity function ($\kevol=1.6\,\pm\,0.1$ for the independent log-Normal \Ep\ and $\kevol=2.1\,\pm\,0.1$ for the intrinsic \EpL\ scenario).
The value of $\kevol=2.1\,\pm\,0.1$ is in agreement with previous studies which find $\kevol=2.1\pm0.6$, \citep{Salvaterra2012} and $\kevol=2.5$, \citep{Pescalli2016} using similar \Ep\ scenarios.
\item  Results obtained with an LGRB rate strictly proportional to the cosmic star formation rate are consistent with results obtained by fixing $\kevol$ and leaving the LGRB rate free to vary. In the intrinsic \EpL\ case, this corresponds to a strong cosmic evolution of the luminosity function ($\kevol=2$) and in the independent log-Normal \Ep\ case, this corresponds to an intermediary case between $\kevol=1$ and $\kevol=2$.
For this reason in the following, the discussion will focus on the scenarios where the LGRB rate is left free to vary.
\item The intrinsic \EpL\ correlation scenarios find an intrinsic slope of the correlation significantly shallower than observed ($\alphaA \simeq 0.3$ versus $\sim 0.5$ observed) and a larger dispersion ($\sigmaEp\simeq 0.45$ versus $\sim 0.28$), close to the best value derived in the independent log-Normal \Ep\ scenario. The value of these best parameters are independent of the value of the scenario for the cosmic evolution of the luminosity and/or the rate.
\item The behaviour of the parameters of the luminosity function and the redshift distribution is similar in both scenarios; as \kevol\ increases, \Lstar\ and the slopes of the redshift distribution, $a$ and $b$, decrease.
Comparing the two scenarios for the peak energy \Ep, we find very similar results regarding the luminosity function for the same value of \kevol: $p$ differs by less than 0.1, \Lstar\ is typically bigger by a factor $\sim 2$ in the intrinsic \EpL\ correlation scenario.
We also find similar results regarding the redshift distribution for the same value of \kevol: the break redshift $z_m \simeq 2.1-2.2$ independently of \kevol\ and of the \Ep\ scenario, the high redshift slope $b$ varies strongly with \kevol\ and is similar in both \Ep\ scenarios and finally the low redshift slope $a$ also varies strongly with \kevol\ but is typically steeper by $+0.2$ in the intrinsic \EpL\ correlation scenario.
\item Finally, the results are comparable for both scenarios regarding the peak energy distribution, which is expected given that \sigmaEp\ is the same and that the correlation slope \alphaA\ is shallow.
\item Interestingly, the local LGRB rate \nGRBo\ increases with \kevol\ in the independent log-Normal \Ep\ scenarios while it stays roughly constant in the intrinsic \EpL\ correlation scenarios.
\end{itemize}
We conclude that we cannot distinguish between an independent log-Normal \Ep\ scenario and an intrinsic \EpL\ correlation scenario solely using the constraints presented in Sect.~\ref{sec:obs_constraints}; we discuss this in more detail in Sect.~\ref{subsec:disc_EpL_correl} using additional cross-checks.
Nonetheless, many results do not depend on the \Ep\ scenario and can be robustly discussed equivalently; this is the case in particular regarding the redshift evolution of the luminosity function and/or of the LGRB efficiency.
Using only the intensity, spectral and redshift constraints we cannot distinguish between luminosity-evolving and efficiency-evolving scenarios.
This degeneracy has been an issue in LGRB population models based on smaller, complete samples of bright LGRBs from {\it Swift} \citep[see e.g.][]{Salvaterra2012,Pescalli2016}.
Despite using fainter and larger samples, we find that lifting this degeneracy remains difficult; we discuss this in more detail using additional cross-checks in Sect.~\ref{subsec:disc_cosmic_evolution}.

\section{Discussion}\label{sec:disc}

\subsection{Is there an intrinsic spectral-energetic correlation?}\label{subsec:disc_EpL_correl}

The observed \EpL\ plane of the extended BAT6 sample is a powerful tool to compare the predictions of the intrinsic \EpL\ correlation and the independent log-Normal \Ep\ scenarios.
There has been significant debate concerning the causes of this observed correlation, whether due to observational selection effects or not \citep[e.g.][]{Nakar2005,Band2005,Butler2007,Ghirlanda2008,Nava2008,Heussaff2013}.
Two caveats should be noted while comparing our population to the observations.
The first is that the observed \eBATsix\ sample is missing about $\sim$17\% of measurements in the \EpL\ plane; the second is that $\sim$25\% of the original BAT6 sample have peak energies only determined by \textit{Swift}/BAT, which can be inaccurate for determining spectral parameters due to its small bandwidth (15-150 keV).
The quantification of these effects on the observed \eBATsix\ \EpL\ plane is beyond the scope of this paper and we do not perform rigorous statistical tests for this reason; this comparison is meant more as an indication.
The observed \EpL\ plane is shown for our mock eBAT6 sample in Fig.~\ref{fig:EpL_all} for our best fit models with the two extreme values $\kevol=0$ and $2$ (no/strong luminosity evolution) and the two scenarios regarding the peak energy \Ep.
Looking at this plane for the independent log-Normal \Ep\ scenarios with mild or no evolution, we find a dearth of high-L bursts compared to the observations (panel (a) of Fig.~\ref{fig:EpL_all}); in the case of strong evolution, this discrepancy is less pronounced (panel (b) of Fig.~\ref{fig:EpL_all}).
On the other hand, the luminosity distributions of the intrinsic \EpL\ correlation scenarios agree well with the observations for all values of \kevol\ (panels (c) and (d) of Fig.~\ref{fig:EpL_all}).
For this reason this scenario is preferred and the following discussions will be focused on this scenario, although we checked that the general trend was the same for the independent log-Normal \Ep\ scenario.
This suggests that the observed \EpL\ correlation cannot be explained only from selection effects, in agreement with \citet{Ghirlanda2008,Nava2008,Ghirlanda2012}.

Nonetheless, looking at the best fit values in Tab.~\ref{tab:pop_result_A}, we find that the slope of the intrinsic relation \alphaA\ is shallower than the observed relation ($\sim$0.3 intrinsic versus 0.54 observed) and the intrinsic scatter is larger ($\sim$0.44 intrinsic versus 0.28 observed).
This implies that some selection effects do shape the observed correlation by making it more pronounced and narrower, as is suggested by the detection limits shown dashed, dot-dashed and dotted for redshift 0.3, 2 and 5 respectively in Fig.~\ref{fig:eBAT6_EpL}.
This result is in good agreement with the current understanding of the GRB prompt emission physics.
The three main possible emission mechanisms 
(internal shocks,  \citealt{Rees1994,Kobayashi1997,Daigne1998} ; magnetic reconnection, \citealt{Usov1994,Spruit2001,Giannios2005,Zhang2011,McKinney2012} ;  
dissipative photosphere, \citealt{Meszaros2000,Beloborodov2017})
indeed predict that the peak energy increases with the luminosity, but depends also on several other parameters, usually linked both to the jet properties (e.g. Lorentz factor) and the microphysics of the emitting region.
Therefore, theoretical models  predict a general tendency for the peak energy to increase with luminosity, but never a strict and narrow \EpL\ correlation unless several physical parameters of the model are strongly correlated.
This has for instance been studied precisely in the case of the internal shock model by \citet{Barraud2005,Mochkovitch2015}.
Our new result can now better specify which intrinsic property of the prompt GRB emission should be compared to theoretical models: a weak correlation with a large $\sim 0.4$ dex scatter of the form
\begin{equation}
\Ep\simeq \left(500-600\, \mathrm{keV}\right)\, \left(\frac{L}{1.6\, 10^{52}\, \mathrm{erg/s}}\right)^{0.3}
\end{equation}

\subsection{The population of soft bursts}\label{subsec:disc_XRFs}
It is remarkable that in all scenarios, the intrinsic distribution of LGRBs (shown as grey contours in Fig.~\ref{fig:EpL_all}) is dominated by largely undetected low-luminosity bursts, which are also mostly low-\Ep\ bursts in the preferred scenario with a moderate intrinsic \EpL\ correlation, in agreement with previous studies \citep{Daigne2006}.
This could be linked to observed subclasses of GRBs such as low-luminosity GRBs (LL-GRBs, \citealt{Liang2007,Virgili2009,Stanway2014}), X-ray Rich GRBs or X-ray Flashes (XRR-GRBs and XRFs respectively, \citealt{Barraud2005,Sakamoto2005,Sakamoto2008}).
With the spectral range and sensitivity of current instruments, their observational samples are still quite small \citep{Virgili2009}, making the precise description of their properties challenging.
For XRFs, the $\gamma$-ray spectra are not very well characterised as the \Ep\ is often below the threshold of detectors and furthermore, the afterglows of XRR/XRFs are rarely observed \citep{Sakamoto2008}.

If the intrinsic spectrum-luminosity correlation is indeed valid, one would expect a link between LL-GRBs and XRR-GRBs/XRFs.
However we note that the observed frequency of these events is larger than the predictions of our model.
For instance, \citet{Sakamoto2005} proposed a classification based on the softness $S$ defined as the ratio of energy fluences in the 2-30~keV and 30-400~keV bands.
Among bursts detected both by the \textit{HETE2}/WXM and FREGATE instruments, they find 22\% of classical GRBs with $S<0.3$, 42\% of XRR-GRBs with $0.3< S< 1$ and 36\% of XRFs with $S<1$.
For the corresponding mock sample (see Tab.~\ref{tab:mock_samples}) in our model, the fraction of XRFs is much lower ($\sim 4\%$) and even XRR-GRBs are underrepresented ($\sim 30\%$).
This suggests that either the luminosity function of LGRBs has a different slope at low luminosity which is unconstrained by our model or that XRFs are not just the low-luminosity tail of the classical LGRB population but rather a full-fledged distinct population of their own.
These two scenarios have different implications in terms of emission mechanisms and progenitor physics and distinguishing between them would require modelling the population of XRFs in details.
Such a dedicated study remains difficult to this day due to the small size of samples of such soft events often without any afterglow or redshift measurements \citep[but see][]{Jenke2016,Katsukura2020}.
We can hope that the situation will improve in the future thanks to a new generation of satellites with a lower energy threshold and good localisation capabilities such as \textit{SVOM}/ECLAIRs (4 - 120~keV, \citealt{Godet2014,Wei2016}) and THESEUS (2~keV - 20~MeV, \citealt{Amati2018}).

\subsection{Cosmic evolution of LGRBs: rate or luminosity?}\label{subsec:disc_cosmic_evolution}

\begin{figure}[!ht]
\centering
\includegraphics[width=\linewidth]{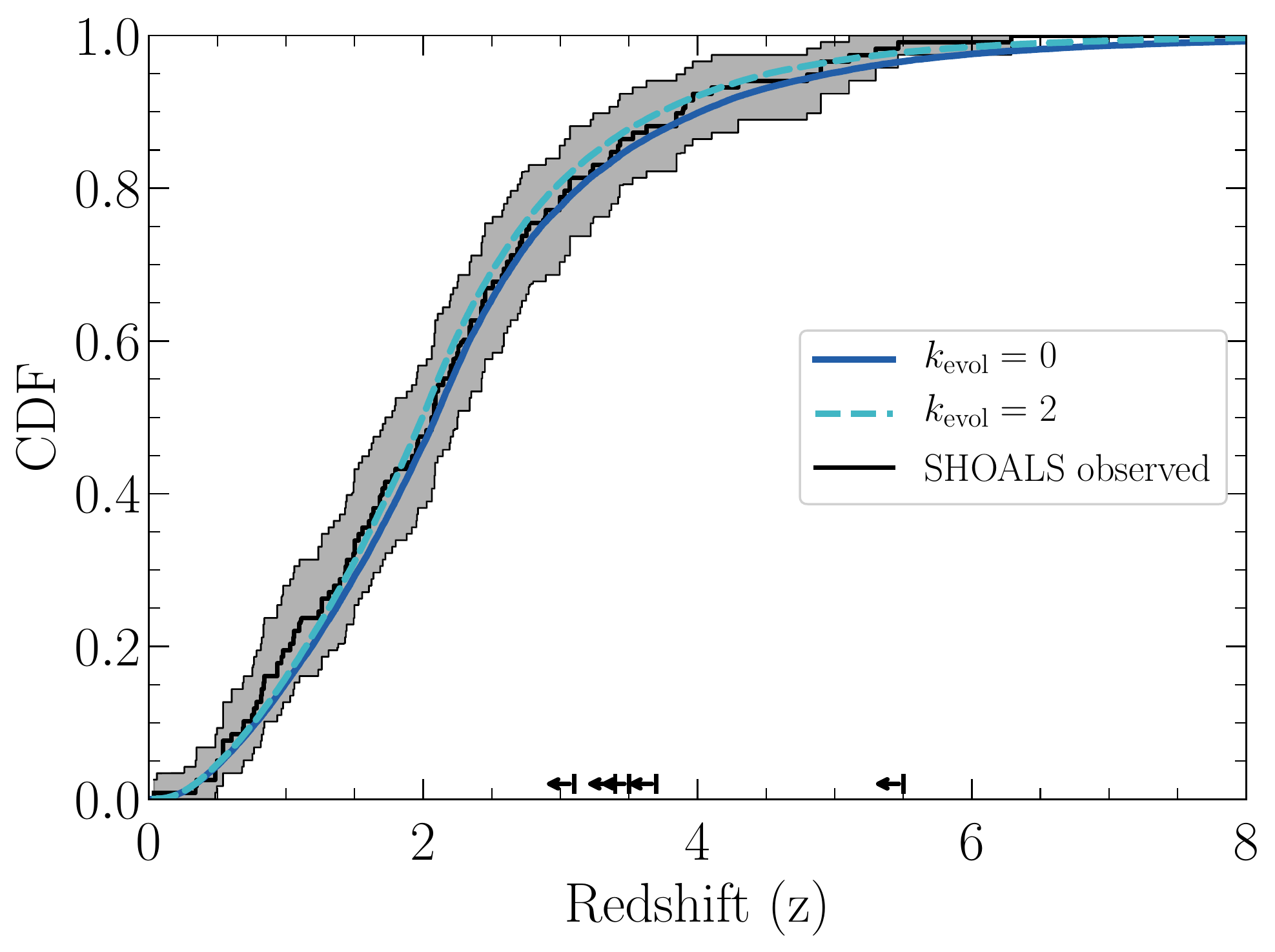}
\caption{Cumulative redshift distribution of the observed SHOALS sample in black; the 95\% confidence bound calculated from bootstrapping is shown in grey and the upper limits are shown as arrows at the bottom of the plot.
The predictions for the intrinsic \EpL\ correlation scenarios with $\kevol=0$ and $\kevol=2$ are shown in blue and dashed light-blue respectively.}
\label{fig:SHOALS_z_distr}
\end{figure}

\begin{figure}[!ht]
\centering
\includegraphics[width=0.48\textwidth]{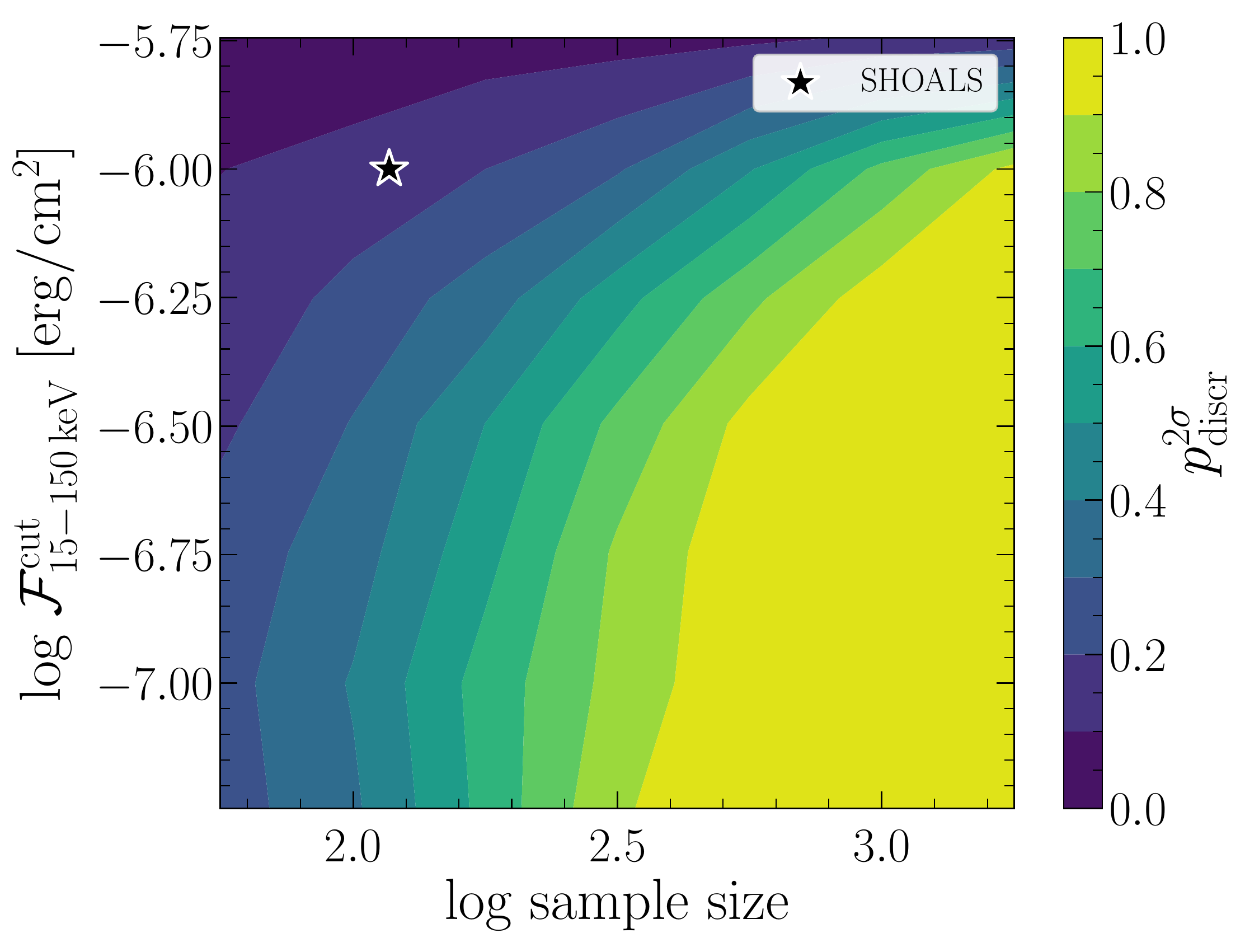}
\caption{Fluence cut - sample size plane for the intrinsic \EpL\ correlation scenario. The colour shading corresponds to the probability of discriminating between the $\kevol=0$ and the $\kevol=2$ scenarios at the $2\,\sigma$ (95\%) confidence level. The star corresponds to the current SHOALS sample of \citet{Perley2016}.
}
\label{fig:flnc_sampsize}
\end{figure}

Having found a number of good models that fit the three observational constraints described in Sect.~\ref{sec:obs_constraints}, we can compare the predictions of each model for the redshift distribution of the SHOALS complete, unbiased sample \citep{Perley2016b,Perley2016} which is $\sim20\%$ larger than the eBAT6 sample.
This comparison is used as a cross-check rather than a constraint since it involves an additional step, namely the calculation of a fluence, which introduces extra uncertainties.
Calculating fluences is not straightforward in our model because we do not simulate a light curve for each LGRB drawn; instead we draw \Cvar\ and \Tnt\obs\ following the prescription presented in Sect.~\ref{subsec:pop_Cvar} and, using the peak photon flux \Npk\ following Eq.~\ref{eq:pflx}, we get the photon fluence $\mathcal{N}$ (in units of [\phcm] in the same energy interval $\left[\Eminobs;\Emaxobs\right]$) with:
\begin{equation}\label{eq:flnc}
\mathcal{N}=\Cvar \, \Tnt \, \Npk
\end{equation}
This assumes that the time-integrated prompt spectrum is the same as the peak-brightness prompt spectrum which is a strong additional assumption.
However, using bursts from the observed \bGBM\ sample (with $\Npksub{50}{300}{keV} \geq 0.9\,\phscm$) we find that the peak energy of these two spectra are tightly correlated: $\log(E_p^{Pk}/E_p^{Ti}) = 0.04\,\pm0.2$.
Additionally, we compared the peak flux - fluence planes of our mock samples with observations and found good agreement suggesting that our method is sound.

The SHOALS sample selection is based on the {\it energy} fluence $\mathcal{F}$ (in units of [\ergcm] between \Eminobs\ and \Emaxobs) which can be consequently calculated from:
\begin{equation}\label{eq:erg_flnc}
\mathcal{F}=\Cvar \, \Tnt \, F^{\rm pk}
\end{equation}
where $F^{\rm pk}$ is the peak energy flux (in units of [\ergscm] in the same energy interval) given by Eq.~\ref{eq:erg_pflx}.
The observed SHOALS sample is defined as having $\mathcal{F}_{15-150\,\keV} > 10^{-6}\,\ergcm$ and favorable observing conditions (see \citealt{Perley2016b} for more details).
We construct our simulated SHOALS sample using this energy fluence threshold and look at the predictions regarding the redshift distribution.
We restrict ourselves to scenarios with $\kevol = 0$ (no evolution) and $\kevol = 2$ (strong evolution) as they are the ones that will exhibit the most differences, the resulting distributions are shown in Fig.~\ref{fig:SHOALS_z_distr}.
All populations that fit the observational constraints described in Sect.~\ref{sec:obs_constraints} predict a mock SHOALS redshift distribution that cannot be distinguished from the observed SHOALS at the 95\% confidence level as is attested by performing K-S tests between the cumulative distributions ($p$-values range from 0.7 to 0.95).

This means that even with the sample size and fluence cut of SHOALS (117 objects, $\mathcal{F}_{15-150\,\keV} > 10^{-6}\,\ergcm$), we cannot distinguish between the scenarios with no redshift evolution of the luminosity function ($\kevol=0$) and scenarios with strong evolution ($\kevol=2$).
We investigated for what values of sample size and fluence cut we could distinguish between the $\kevol=0$ and the $\kevol=2$ scenarios using KS-tests; the results are shown in Fig.~\ref{fig:flnc_sampsize} while the detailed methodology is shown in App.~\ref{app:flnc_cut_samp_size_plane}.
This plot illustrates how difficult it is to lift the degeneracy between the cosmic evolution of the LGRB rate and of the LGRB luminosity.
The sample size should be increased by a factor larger than $10$ at fixed fluence cut in order to have a good chance to distinguish between the two types of cosmic evolution at the 95\% confidence level; decreasing only the fluence cut does not really improve the test.
In practice, a combination of lower fluence cut and larger sample size could be used, but even then a factor of $\sim 3$ in both directions would be needed to get a decent probability of discriminating between the two types of cosmic evolution at the 95\% confidence level.
Lifting the degeneracy between luminosity and efficiency cosmic evolution remains difficult using the redshift distribution of reasonably sized samples.

\begin{figure*}
\centering
\includegraphics[width=\linewidth]{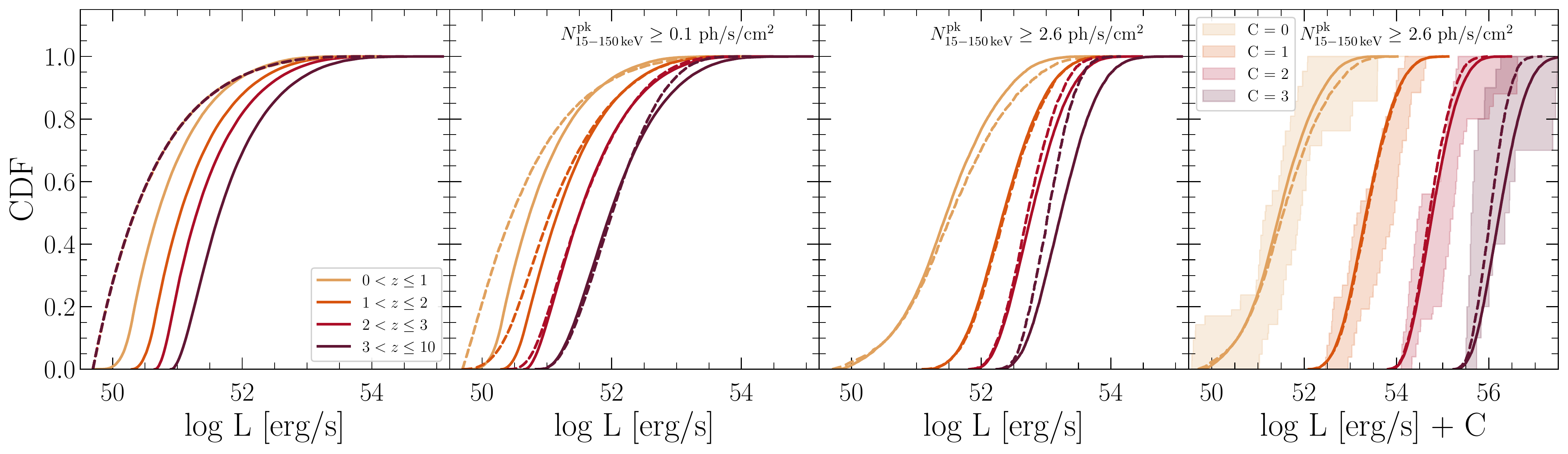}
\caption{Luminosity cumulative distribution function in four different redshift bins for the intrinsic \EpL\ correlation scenario.
The full lines correspond to $\kevol=2$ and the dashed lines to $\kevol=0$.
The leftmost panel shows the entire intrinsic LGRB population, while the center left and center right panels show bursts with $\Npksub{15}{150}{keV}\geq 0.1~\phscm$ and $2.6~\phscm$ respectively.
The rightmost panel is the same as the center right panel but adding the observed eBAT6 sample and where each curve was shifted by a constant $C$ for clarity.
The shaded area represents the 95\% confidence region for the CDF of the observed eBAT6 sample, calculated using bootstrapping (without taking into account limits).}
\label{fig:L_CDF_zbins}
\end{figure*}

Another approach to lifting the degeneracy is to look directly at the luminosity distribution in different redshift bins.
Fig.~\ref{fig:L_CDF_zbins} shows the luminosity CDF in four redshift bins for the intrinsic \EpL\ correlation scenario at different peak flux cuts.
The scenario with $\kevol=0$ is shown in dashed while $\kevol=2$ is shown in full.
As expected, if one had access to the entire population (leftmost panel), it would be easy to distinguish between no luminosity evolution and strong luminosity evolution since the CDF would be the same in all redshift bins for the case $\kevol=0$.
The difficulty resides in the fact that as we increase the peak flux cut (as shown in middle panel of Fig.~\ref{fig:L_CDF_zbins}), we remove mostly lower luminosity bursts and shift the distributions towards higher luminosities, essentially creating the same effect as luminosity evolution.
This is illustrated in the center left and center right panels of Fig.~\ref{fig:L_CDF_zbins} where the same plots are shown but for bursts with $\Npksub{15}{150}{keV}\geq 0.1~\phscm$ and $2.6~\phscm$ respectively.
The rightmost panel of Fig.~\ref{fig:L_CDF_zbins} shows the same curves as in the center right panel but adding the 95\% confidence area from the observed eBAT6 sample.
In addition to the good agreement between our models and the observed data, we can see with current sample sizes ($\sim 100$ objects) that it is not possible to argue in favor of a specific value for \kevol.
Both of these tests become even more difficult if we include the intermediate scenarios with $\kevol=0.5$ and $\kevol=1$.

\subsection{The LGRB rate and its connection to the cosmic star formation rate}\label{subsec:disc_LGRB_SFR}

\subsubsection{The LGRB production efficiency by massive stars}

\begin{figure}
\centering
\includegraphics[width=\linewidth]{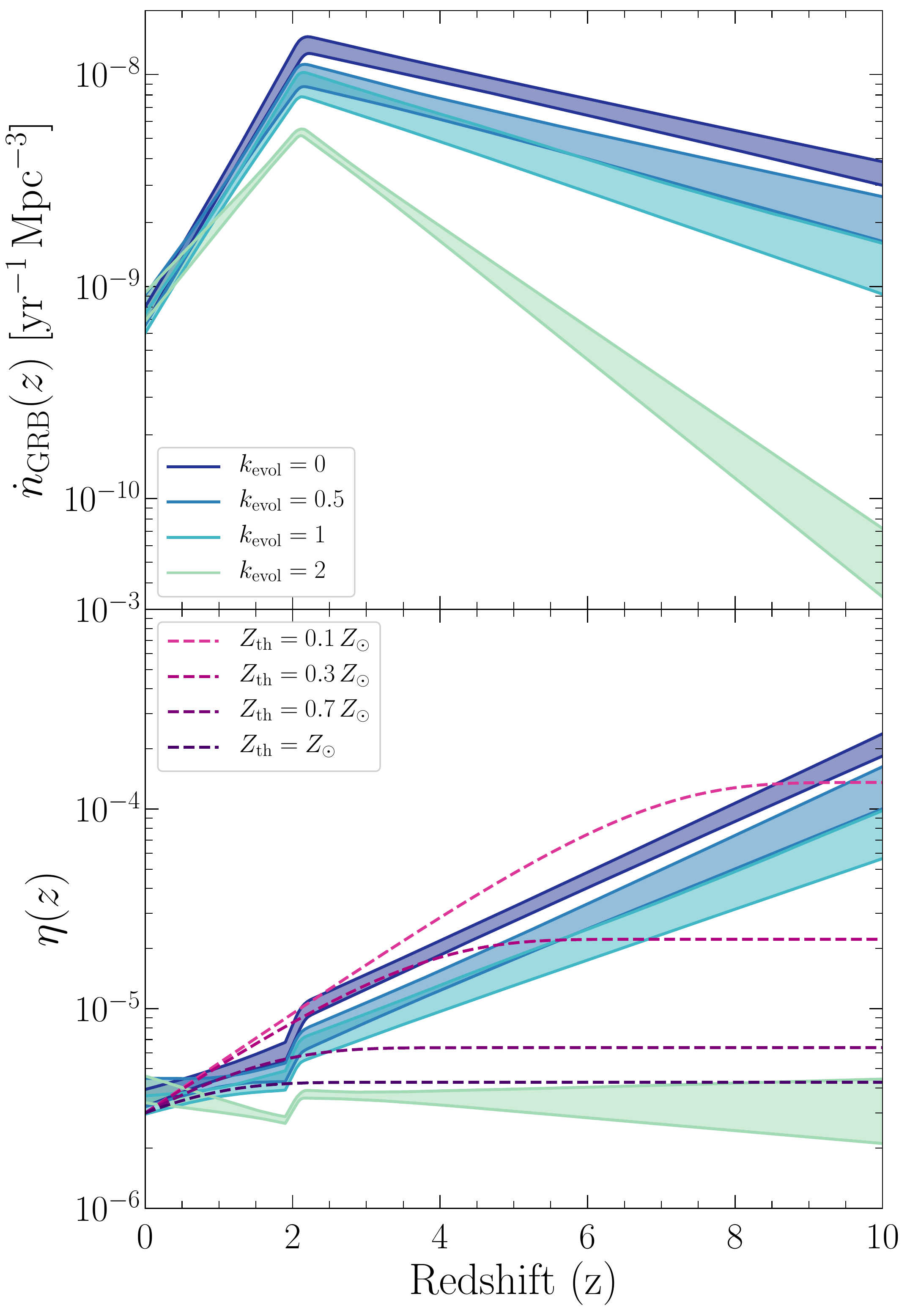}
\caption{{\it Top panel:} comoving LGRB rate density as a function of redshift for the best fitting models of the intrinsic \EpL\ correlation scenarios.
{\it Bottom panel:} LGRB efficiency $\eta(z)$ as as function of redshift obtained with Eq.~\ref{eq:LGRB_eff}. The estimation of the fraction of star formation below $Z_{\rm th}$ from Eq.~5 of \citet{Langer2006} is shown, rescaled for comparison, in dashed.
The shaded area represents the 95\% confidence interval.}
\label{fig:LGRB_efficiency}
\end{figure}

The top panel of Fig.~\ref{fig:LGRB_efficiency} shows the LGRB comoving rate density as a function of redshift for the intrinsic LGRB population in the intrinsic \EpL\ correlation scenario for different values of \kevol.
The case of $\kevol=2$ follows closely the star formation rate density, while the high redshift slope gets shallower as \kevol\ decreases.
Models with lower values of \kevol\ have higher global rates of LGRBs.
Using Eq.~\ref{eq:LGRB_eff}, we can derive the LGRB efficiency as a function of redshift, shown in the bottom panel of Fig.~\ref{fig:LGRB_efficiency}.
Assuming $p_{cc}$ and $\bar{m}$ are constant with $z$, models with $\kevol < 2$ imply some amount of evolution of the LGRB efficiency with redshift, with the strongest evolution found for models with $\kevol=0$.
There are many possible causes for this increasing LGRB efficiency with redshift, one of which is metallicity.
This is in line with observational studies of complete, unbiased samples of LGRB host galaxies \citep{Hjorth2012,Vergani2015,Japelj2016,Perley2016b,Vergani2017,Palmerio2019} which find that metallicity is the driving factor behind the LGRB production efficiency.
The fraction of star-formation happening below a given metallicity (from Eq.~5 of \citealt{Langer2006}) is shown in the bottom panel of Fig.~\ref{fig:LGRB_efficiency} as dashed lines, arbitrarily scaled.
We can see some similarity between our derived efficiency and these curves, although the behaviour at high redshift is different.
This might be an indication of an additional break at $z\geq6$ in the redshift distribution of LGRBs, although with the amount of data available at these redshifts to date, it would be hard to constrain.
Another possibility is that the effect of metallicity is more complicated than a simple threshold.
Indeed, metallicity can also play an indirect role by influencing the stellar IMF \citep[e.g.][]{LaBarbera2013} or the fraction of binary progenitors \citep{Chrimes2020}.

\subsubsection{The Local LGRB rate}

The local LGRB rate density for LGRBs pointing towards us we derived with our population model is  $\nGRBo = 0.7-0.8~\yrGpc$ for the intrinsic \EpL\ correlation scenarios and $\nGRBo = 1.0-1.5~\yrGpc$ for the independent log-Normal \Ep\ scenarios (see Tab.~\ref{tab:pop_result_LN} and \ref{tab:pop_result_A} for details).
This local LGRB rate density is highly dependent on the value of $\Lmin=5\times10^{49}~\ergs$; the following equation can be used to extrapolate to lower values of \Lmin\ for a Schechter or power-law function assuming that the additional LGRBs are not detected in the observed samples:
\begin{equation}
\nGRBo(\Lmin) = \nGRBo(L_{\rm min}^{\rm ref})\,\left( \frac{\Lmin}{L_{\rm min}^{\rm ref}} \right)^{1-p}
\end{equation}
where $p$ is the slope of the luminosity function and $L_{\rm min}^{\rm ref}=5\times10^{49}\ergs$ in our case.
Taking into account the difference in \Lmin, our values are mostly consistent with those of previous authors (but see App.~\ref{app:other_models}).
We should note however that the slope at low-luminosities may be different (see discussion in Sect.~\ref{subsec:disc_XRFs}) and therefore any extrapolation to lower \Lmin\ should keep this in mind.

Using Eq.~\ref{eq:ndot_GRB} we find $\eta_0 =\left( 3-6\right)\times10^{-6}$,
as shown in the bottom panel of Fig.~\ref{fig:LGRB_efficiency}, or in other words $3-6$ LGRBs pointing towards us for every million core-collapse.
This can be extended by dividing by the average jet opening angle to estimate the total number of LGRBs occurring per core-collapse (including LGRBs pointing away from us).
Using $\langle \theta_{\rm jet} \rangle \simeq 5\degree$ \citep{Harrison1999,Ghirlanda2013,Gehrels2013,LloydRonning2020}, we get $0.8-1.6$ LGRBs pointing in any direction per thousand core-collapse at $z=0$ and up to a factor 10 higher at $z=6$ (depending on the value of \kevol).
This highlights the fact that LGRBs are rare events among transients, requiring specific conditions to form and that these conditions may be more readily met earlier in the Universe.

\subsubsection{Measuring the CSFRD with LGRBs ?}
Due to their association with massive stars, LGRBs have been used as probes of the cosmic star formation rate density \citep{Kistler2008,Robertson2012,Hao2020}, in particular up to high redshift where estimations from rest-frame UV measurements are plagued by dust uncertainties \citep{Bouwens2015,Bouwens2016}.
However, in order to estimate the CSFRD from the LGRB rate, one must make hypotheses on the redshift evolution (or lack thereof) of the LGRB luminosity function and the LGRB production efficiency.
Typically what is done is to calibrate the relationship between the CSFRD and the rate of bright LGRBs (by using only LGRBs above a certain luminosity) at low redshift ($z<4$) and extrapolate this relationship out to high redshift.
In doing so, authors usually assume a non-evolving luminosity function, however as we have shown in Sect.~\ref{subsec:disc_cosmic_evolution}, there is a strong degeneracy between the cosmic evolution of the LGRB efficiency and the LGRB luminosity function which cannot be lifted with the current sample sizes.
This means that estimates of the CSFRD using LGRBs should take into account the uncertainty due to the possible redshift evolution of the luminosity function and reflect this in their error bars.
This correction is not a small effect, looking at Fig.~\ref{fig:LGRB_efficiency}, we see that at $z=6$, it amounts to a factor of at least 10 between the extreme scenarios.

\subsection{Strategy for high redshift detection}\label{subsec:disc_highz_strat}
\begin{figure*}
\centering
\includegraphics[width=\linewidth]{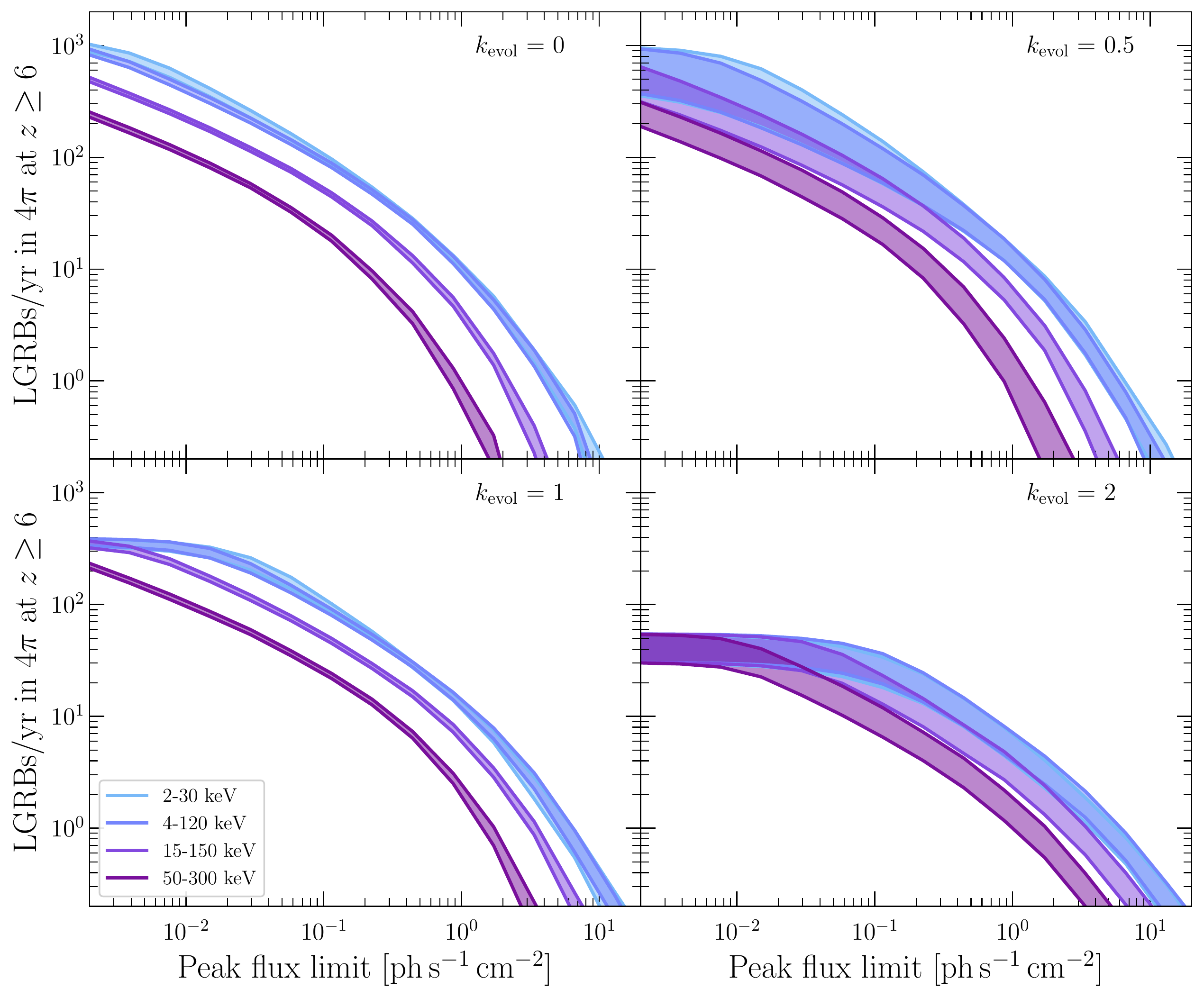}
\caption{All-sky rate of LGRBs at $z\geq6$, as a function of limiting peak flux for different energy bands in the intrinsic \EpL\ correlation scenario.
Each panel corresponds to a luminosity evolution scenario (from no evolution at $\kevol=0$ to strong evolution at $\kevol=2$).
To obtain the actual observed rate for a given mission, one should take into account the field of view of the instrument and its duty cycle.
The shaded area represents the 95\% confidence interval.}
\label{fig:R_highz}
\end{figure*}

Using our intrinsic LGRB populations, we can investigate which type of strategies are most effective at observing high redshift LGRBs, a primary goal for using the full potential of LGRBs as a cosmological tool.
In Fig.~\ref{fig:R_highz} we show the expected all-sky yearly rate of LGRBs at $z\geq6$ as a function of limiting peak flux for different energy bands.
We see that, at a given peak flux limit, the rate of high redshift LGRBs is higher for softer energy bands, and depends in particular on the lower energy threshold of the band, in line with the results of \citet{Ghirlanda2015}.
This bodes well for the SVOM mission to be launched in 2022 \citep{Wei2016}, whose coded-mask telescope ECLAIRs is designed to trigger at these lower energies with a low-energy threshold of 4~keV \citep{Godet2014}, and even more for the THESEUS\footnote{\url{https://www.isdc.unige.ch/theseus/}} mission in discussion for the next decade, whose XGIS instrument will detect GRBs in the 2~keV - 20~MeV energy band with a larger effective area \citep{Amati2018}. 
Furthermore, we see that for models with $\kevol=2$, the rate of LGRBs at $z\geq6$ reaches a plateau at a peak flux limit $\Npk\sim10^{-2}~\phscm$; this implies that the whole of the population is seen and going to lower peak fluxes will not increase the proportion of high redshift LGRBs.
This could be an interesting test to distinguish between different luminosity evolution scenarios.
However, the possibility of getting unbiased, redshift-complete samples of LGRBs down to such low peak fluxes remains to this day unlikely. 

\section{Conclusion}\label{sec:concl}
We created a model for the intrinsic population of LGRBs which we constrained by comparing to carefully selected observational constraints, specifically tailored to constrain all major aspects of the population.
We explored two main scenarios concerning the distribution of the peak energy \Ep: one where the peak energy follows a log-Normal distribution independently of other properties (independent log-Normal \Ep) and one where the peak energy is correlated to the luminosity (intrinsic \EpL\ correlation).
In addition, in each case we explored four scenarios for the evolution of the luminosity function: no evolution ($\kevol=0$), mild evolution ($\kevol=0.5$), evolution ($\kevol=1$) and strong evolution ($\kevol=2$).
We derived the best fit parameters in each scenario using a Bayesian Markov Chain Monte Carlo exploration scheme with an indirect likelihood.
Using two additional cross-checks we discussed the results of our intrinsic population in the context of the intrinsic spectral-energetics correlations, soft bursts, the cosmic evolution of the luminosity function, the LGRB rate and its connection to the cosmic star formation rate and the strategy for detection of high redshift bursts.
Our conclusions can be summarised as follows:

\begin{itemize}
    \item The observed \EpL\ correlation cannot be explained only by selection effects, however these do play a role in shaping the relation. Compared to the observed relation, we derive a shallower slope for the intrinsic correlation $\alphaA\sim0.3$ and a larger scatter of $\sim0.4~\rm dex$.
    \item Our intrinsic population is not able to accurately describe the proportion of soft bursts (e.g. XRR-GRBs, XRFs) which are expected to be weak following the spectral-energetics correlation; this may imply a different slope of the luminosity function at low luminosities or a separate population all-together, either case is difficult to constrain with current samples.
    \item We confirm the degeneracy between cosmic evolution of the LGRB rate and of the luminosity function.
    Under a strong cosmic evolution of the luminosity function ($\kevol=2$) we find that the LGRB rate follows the cosmic star formation rate; however under a non-evolving luminosity function we find a increasingly higher LGRB production efficiency from stars at high redshift.
    \item We show that this degeneracy cannot be lifted with current sample sizes and completeness.
    Compared to the current SHOALS sample, a factor of 3 larger sample size and lower fluence cut would be needed to distinguish between the two most extreme scenarios.
    \item Assuming that the cosmic star formation rate is well measured at high redshift, the scenario with a non-evolving luminosity function ($\kevol=0$) implies an LGRB production efficiency from stars a factor of $\sim10$ higher at $z=6$ than at $z=0$; assuming it is not, this could be instead interpreted as evidence for the underestimation of the cosmic star formation rate at high redshift.
    \item The degeneracy between luminosity and redshift evolution is usually not taken into account in studies aiming at measuring high redshift star formation rate using LGRBs, we emphasise the need to do so as it impacts the estimation of the LGRB rate by up to a factor of 10 at $z=6$ and 50 at $z=9$.
    \item We derive a local LGRB rate density for LGRBs pointing towards us $\nGRBo \sim 0.8~\yrGpc$ in the preferred intrinsic \EpL\ scenario which translates to $\sim1$ LGRB pointing in any direction per thousand core-collapse at $z=0$, assuming a mean jet opening angle of $\langle \theta_{\rm jet} \rangle \simeq 5\degree$. 
    \item We confirm that the best strategy for detecting high redshift bursts is to use detectors with a soft low-energy threshold of $\sim$ a few keV such as the ECLAIRs telescope on-board the \textit{SVOM} satellite to be launched in 2022 or the XGIS instrument of the future THESEUS mission.
    We find that the combination of such a low energy threshold with a good sensitivity could help to partially lift the degeneracy between luminosity and rate evolution.
\end{itemize}

\section*{Acknowledgements}
This work has been partially supported by the French Space Agency (CNES).
JTP would like to thank Sébastien Carassou, Tom Charnock and Jens-Kristian Krogager for invaluable discussion regarding the statistical framework used in this paper.
This research has made use of Astropy, a community-developed core Python package for Astronomy (Astropy Collaboration, 2013).
The corner plots used in this paper for representing multi-dimensional datasets were performed with the \textsc{python} corner.py code \citep{corner.py}.

\bibliographystyle{aa} 
\bibliography{biblio}

\appendix
\section{Band function}\label{app:band}
\begin{figure*}
\begin{center}
\begin{subfigure}{.50\textwidth}
    \centering
    \includegraphics[width=\linewidth]{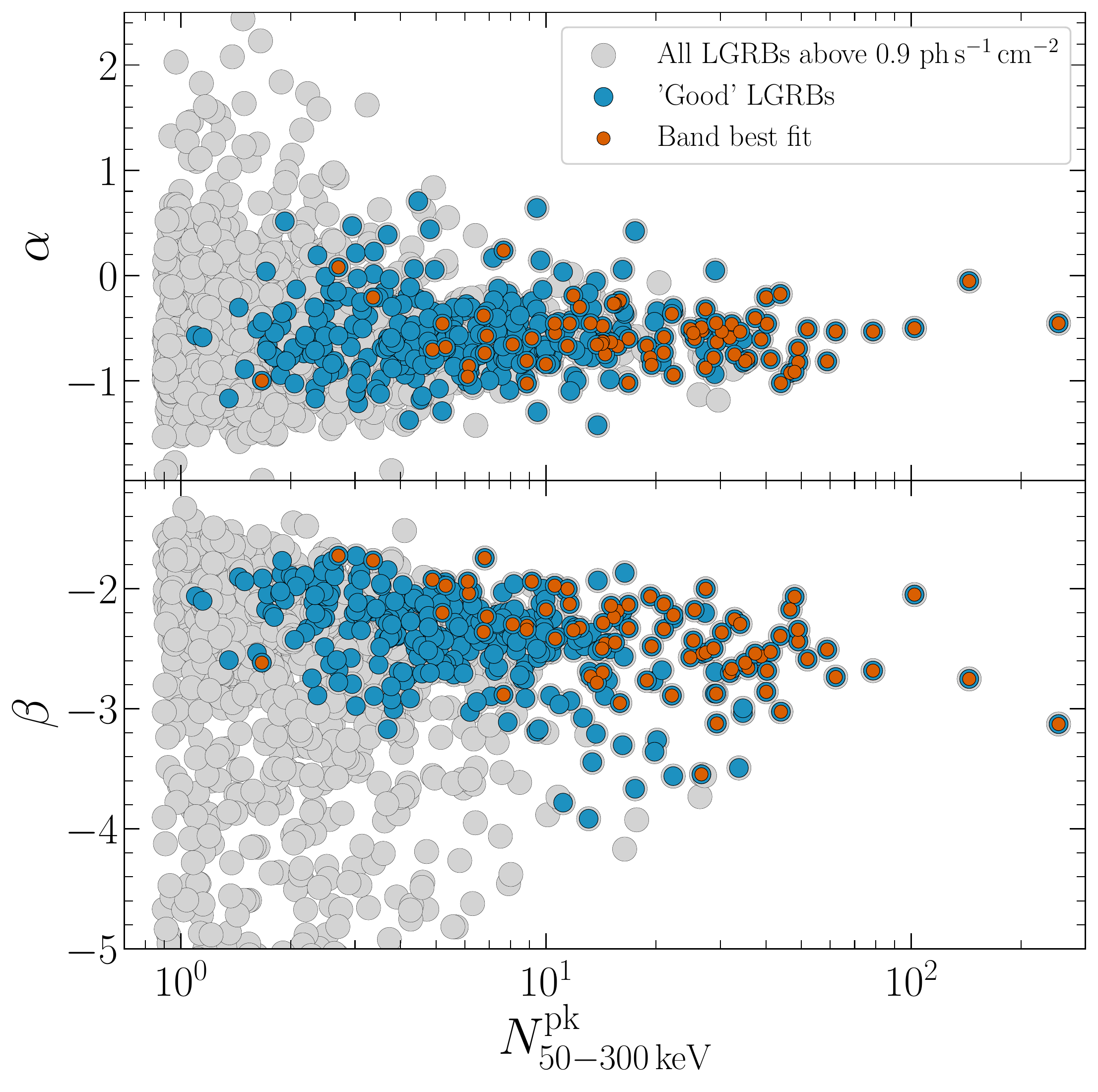}
\end{subfigure}
\begin{subfigure}{.36\textwidth}
    \centering
    \includegraphics[width=\linewidth]{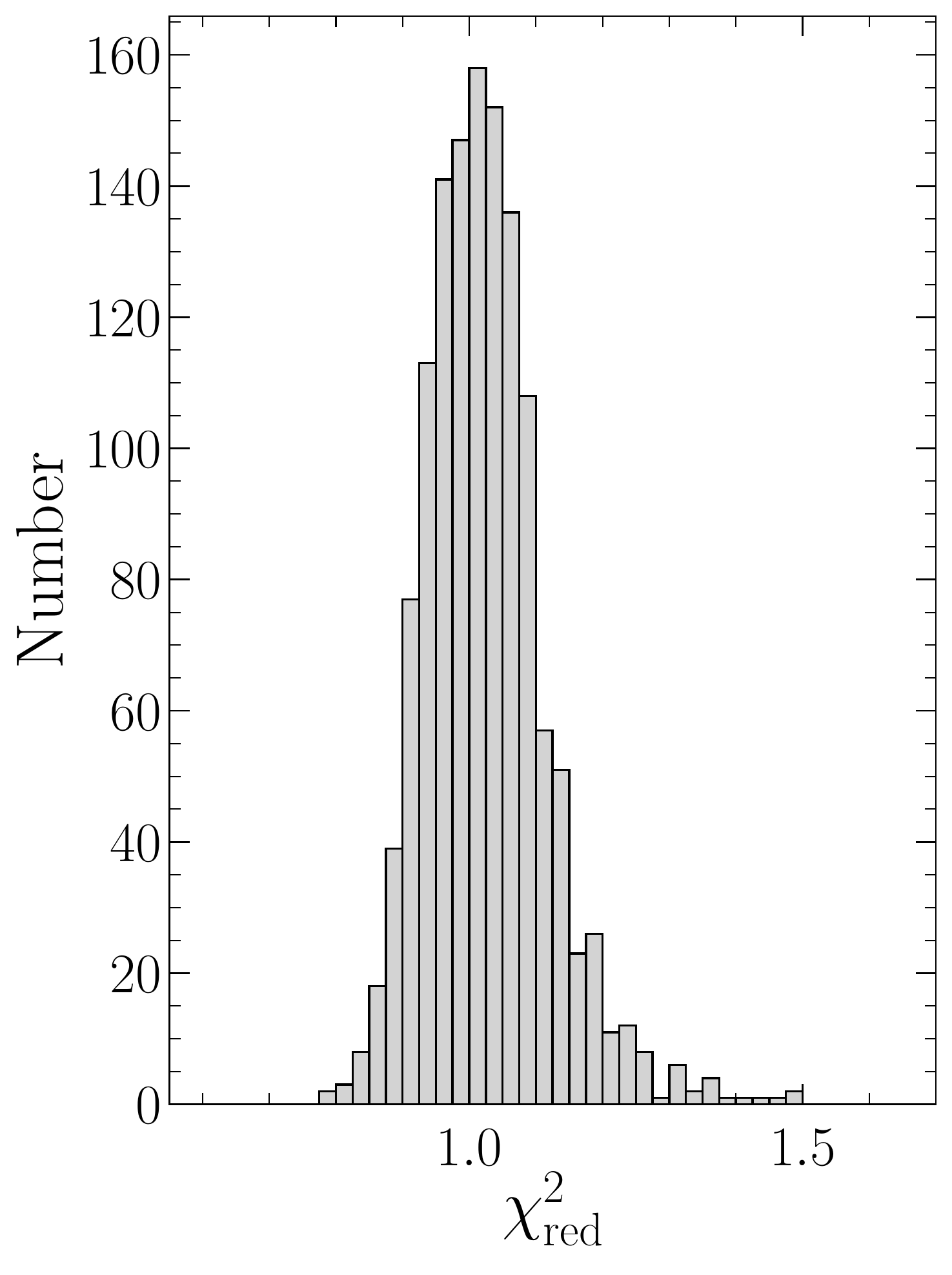}
\end{subfigure}
\caption{Quality of Band fits to GRB spectra from GBM catalogue \citep{Bhat2016}.
{\it Left:} Spectral slopes $\alpha$ (top) and $\beta$ (bottom) as a function of peak flux for the GBM spectral catalogue \citep{Bhat2016}.
The entire catalogue of long GRBs above 0.9 \phscm\ is in light-grey, the bursts complying with the "good" criteria (i.e. small errors on the parameters) are shown in blue, the bursts for which Band is the best-fit spectral model are shown in orange.
{\it Right:} reduced \Chisq\ distribution for Band models from the GBM spectral catalogue \citep{Bhat2016}.}
\label{fig:Band_quality}
\end{center}
\end{figure*}

The Band function \citep{Band1993} is given by:
\begin{equation}
f(E)  = A
\begin{cases}
\left(\frac{E}{100\,\rm{keV}}\right)^{\alpha}\,\exp\left( -(\alpha + 2)\,\frac{E}{E_p} \right) & E \leq E_c , \\
\left(\frac{E}{100\,\rm{keV}}\right)^{\beta}\,\left(\frac{E_c}{100\,\rm{keV}} \right)\,\exp(\beta - \alpha) & E > E_c ,
\end{cases} 
\end{equation}
where $\alpha$ is the low energy slope, $\beta$ the high energy slope, \Ep\ the peak of the $E\,L_{E}$ spectrum, $E_c = \frac{\alpha-\beta}{\alpha+2}\,\Ep$ and $A$ is a normalisation that satisfies $1 = \int_{0}^{\infty}\, f(E)\, \dd E$.
The goodness of fit of this functional form, shown in the right panel of Fig.~\ref{fig:Band_quality} for the 3rd GBM catalogue above 0.9\,\phscm, illustrates that even when Band is not the best fitting model it still provides a good fit to the data (i.e. the reduced \Chisq\ is around 1).

\section{Statistical framework}\label{app:stats}

\renewcommand{\arraystretch}{1.2}

\subsection{Likelihood}
We decided to follow a methodology already used for galaxy catalogs from image extraction \citep{Carassou2017} which is based on a Parametric Bayesian Indirect Likelihood (pBIL, \citealt{Drovandi2015}).
The basic idea is to use an auxiliary likelihood when a complex problem renders its own likelihood intractable.
To this end, we used the binned maximum likelihood method \citep{Barlow1993} which assumes the number of objects in each bin follows a Poissonian law and we construct our auxiliary likelihood such that for each bin $i$, the probability of $o_i$ given the model $s_i$ is:
\begin{equation}
l_i = \frac{e^{-s_i}\,s_i^{o_i}}{o_i!}
\end{equation}
The likelihood for the entire histogram becomes:
\begin{equation}
\mathcal{L} = \prod_{i=1}^b\, \frac{e^{-s_i}\,s_i^{o_i}}{o_i!}
\end{equation}
where $b$ is the total number of bins in the histogram, $o_i$ and $s_i$ are the observed and simulated number count in bin $i$ respectively.
The log-likelihood is thus:
\begin{equation}\label{eq:likelihood_def}
\ln\mathcal{L} = \sum_{i=1}^b\,o_i\ln(s_i)-s_i
\end{equation}
where the factor $\ln(o_i!)$ is neglected since it is a constant and our goal is to maximize \lnL.
This likelihood presents a problem if a single simulated bin is empty therefore we added an infinitesimal $\epsilon=10^{-3}$.
We checked the impact of different values of $\epsilon$ which affected mostly models with bad likelihood (i.e. models with many empty bins); the effect on good models was negligible.
Multiple histograms can be included by simply adding the logarithm of the likelihood (i.e.  multi-plying the likelihoods) however, the importance of a given histogram on the overall likelihood relies on its number of objects.
For this reason, the intensity constraint (see~\ref{subsec:Stern}) has the strongest weight of our constraints due to its large sample size ($N\sim7000$ once the efficiency correction is taken into account).
In order to strengthen the impact of the redshift constraint (weak due to its small sample size, $N=82$), we added a weight of 10 to its log-likelihood.
We tested different weights and this one was chosen as a balance between having a notable impact and being unrealistically constraining with respect to the other constraints.

\subsection{Parameter space exploration}
We decided to use a MCMC scheme to explore the parameter space, with a modified version of the Metropolis-Hastings algorithm called simulated annealing \citep{Kirkpatrick1983}.
This modification is based on the idea of cooling metals and allows for the Markov chain to initally accept worse likelihoods with a factor $\tau=\tau_0$ that decreases at a user-defined rate. 
As the \textit{effective temperature} $\tau$ decreases, the chain is less and less likely to accept worse jumps until $\tau \to 1$, where we return to a classic Metropolis-Hastings algorithm.
This modification to the original Metropolis-Hastings algorithm verifies the condition of ergodicity: regardless of the starting point the Markov Chain will converge to the same stationary distribution.

\begin{figure}
\begin{center}
\includegraphics[width=\linewidth]{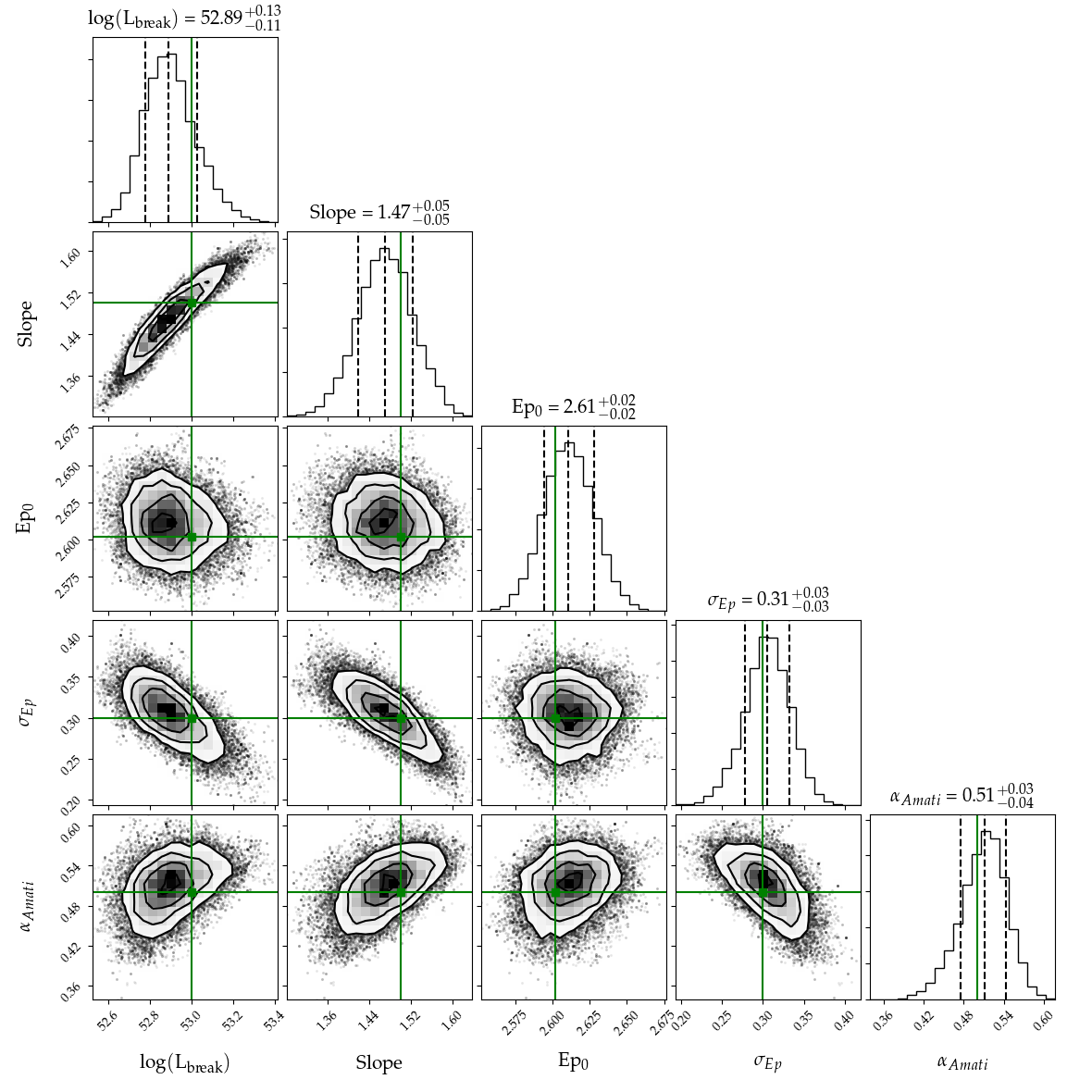}
\end{center}
\caption{Corner plot from the MCMC exploration of fake observations generated from known inputs.
The known true values are shown in green.
}
\label{fig:cp_method_check}
\end{figure}

We also tested the methodological soundness of our algorithm by applying it to data generated by known inputs.
We created mock observations for the intensity, spectrum and redshift constraint which we then used as constraints for our MCMC scheme; the results of the parameter space exploration are presented in Fig.~\ref{fig:cp_method_check}.
We can see the code is able to satisfactorily recover the input parameter values: Schechter luminosity function \{$\Lstar=10^{53}~\ergs$, $p=1.5$\}, redshift distribution following the shape of the cosmic star formation rate density; intrinsic \EpL\ correlation scenario \{$\Epo=10^{2.60}~\keV$, $\sigmaEp=0.30$, $\alphaA=0.50$\}.
\begin{table}[!hb]
\centering
\begin{adjustbox}{max width=\linewidth}
\begin{tabular}{cccc}
\toprule
Functional form    & Parameter& Prior range & Units       \\ \toprule
\multicolumn{4}{c}{Luminosity Function}  \\  \cmidrule(lr){1-4}
\multirow{4}{*}{Schechter Function} & \Lmin  & $5\times10^{49}$   & \ergs       \\
                   & \Lstar  & [$10^{52}$ - $10^{56}$]   & \ergs       \\
                   & slope    & [0.3 - 3]   &             \\ 
                   & \kevol   & [0, 0.5, 1, 2] or [$-3$ - 5] &             \\ \midrule
\multicolumn{4}{c}{Redshift Distribution}  \\ \cmidrule(lr){1-4}
\multirow{3}{*}{Broken Exponential} & a     & [$-3$ - 3] &             \\
                   & b        & [$-3$ - 3] &             \\ 
                   & z$_m$    & [0 - 10] &             \\ \midrule
\multicolumn{4}{c}{Peak Energy Distribution} \\ \cmidrule(lr){1-4}
\multirow{2}{*}{Log-Normal} & \Epo     & [$10^1$ - $10^4$] & keV      \\ 
                   & \sigmaEp & 0.45 &             \\ \cmidrule(lr){2-4}
\multirow{3}{*}{Intrinsic Correlation}& \Epo     & [$10^1$ - $10^4$] & keV      \\
                   & \sigmaEp & [0 - 1.5]  &             \\
                   & \alphaA  & [$-1$ - 1.5] &             \\ \bottomrule
\end{tabular}
\end{adjustbox}
\caption{A summary of the flat prior bounds used for the parameters of our population model. \Lmin\ and \sigmaEp\ were fixed during the parameter exploration and thus have no bounds.}
\label{tab:priors}
\end{table}
\renewcommand{\arraystretch}{1}

\section{Fluence cut - sample size plane}\label{app:flnc_cut_samp_size_plane}
The sample size and fluence cut of SHOALS do not allow us to discriminate between non-evolving ($\kevol=0$) and strongly evolving ($\kevol=2$) scenarios for the luminosity function of LGRBs.
In this section we detail our methodology for determine the probability of discriminating between these two scenarios given a fluence cut and a sample size.

\begin{figure}
\begin{center}
\includegraphics[width=\linewidth]{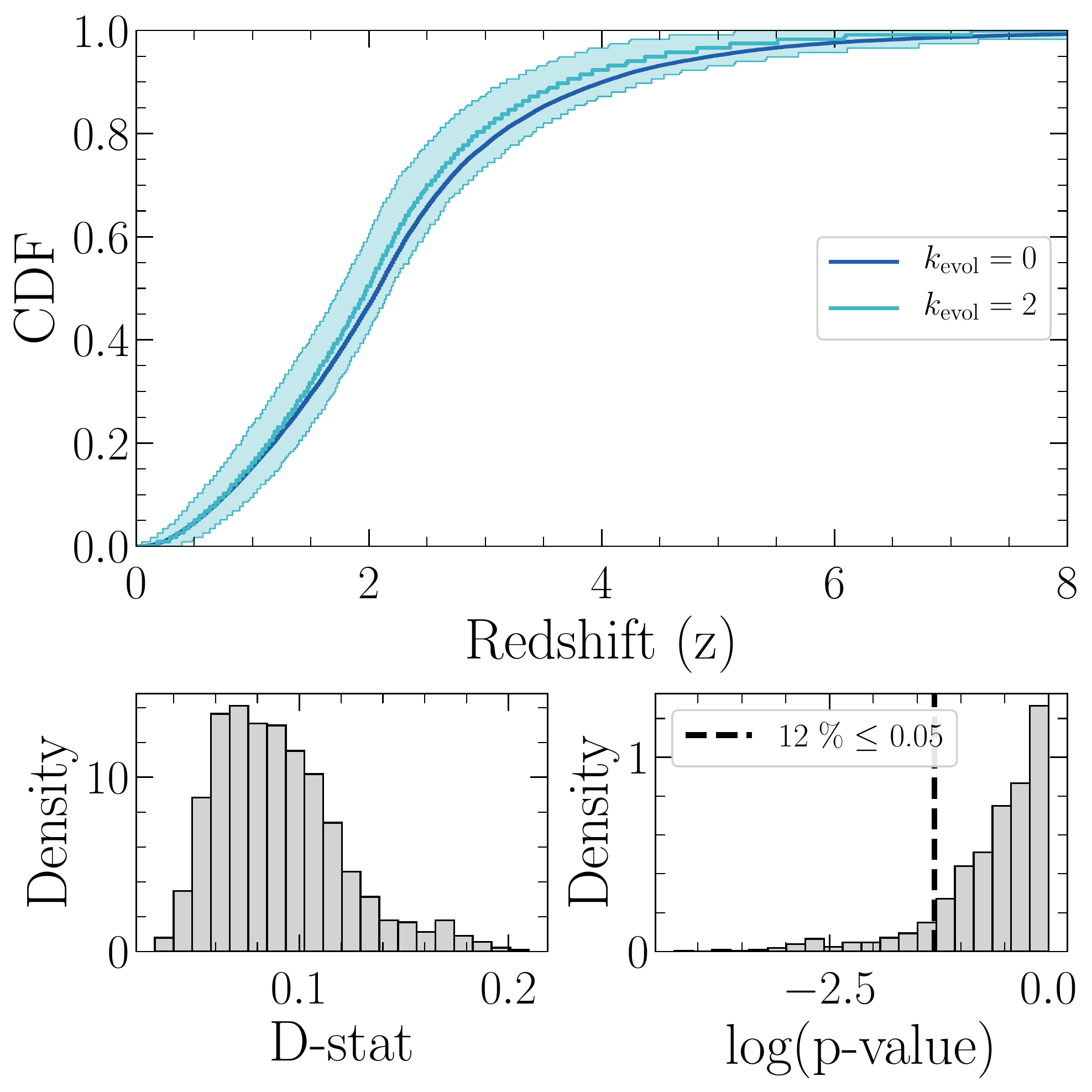}
\end{center}
\caption{{\it Top panel}: Redshift cumulative distribution function of the intrinsic \EpL\ correlation scenarios cut at $\mathcal{F}_{\rm cut}=10^{-6}\,\phcm$ for $\kevol=0$ in dark blue and $\kevol=2$ in light blue.
The $\kevol=2$ scenario is sub-sampled at $N_{\rm sub}=117$; the 95\% confidence bound is calculated from bootstrapping.
{\it Bottom panel}: Distribution of p-values (right) and D-statistics (left) from the K-S tests performed on the bootstrapped samples.
The back dashed vertical line shows a p-value of 0.05, below which it is possible to distinguish between the two distributions at the 95\% confidence level.
}
\label{fig:flnc_sampsize_method}
\end{figure}

First, we cut both scenarios at a given fluence cut $\mathcal{F}_{\rm cut}$.
We then create a subsample from one scenario of size $N_{\rm sub}$ and compute the K-S test between the redshift distributions of this subsample and the sample of the other scenario.
This K-S test is repeated for a number of bootstrap samples $N_{\rm bs}$; the result is a distribution of p-values and D-statistics shown in the bottom panels of Fig.~\ref{fig:flnc_sampsize_method}.
We define the probability of discriminating between the two scenarios at the 95\% confidence level as the fraction of realisations with a p-value below 0.05; in the example of Fig.~\ref{fig:flnc_sampsize_method}, this probability would be 12\%.
We calculate this fraction for different values of $N_{\rm sub}$ and $\mathcal{F}_{\rm cut}$ and draw the contours to obtain the fluence cut - sample size plane shown in Fig.~\ref{fig:flnc_sampsize}.

\section{Comparison with other models}\label{app:other_models}
For each model below, we generated the synthetic population with our own Monte Carlo code using $\NGRB=10^{6}$ LGRBs, assuming the same cosmology and the same functional forms for the intrinsic distributions (rate, luminosity function, etc.) as reported by the authors, and adopting the best fit parameters we found in the papers. We then confront the properties of the resulting population to the observational constraints used in the present study.

\subsection*{Daigne et al. 2006}
\begin{figure*}
\begin{center}
\includegraphics[width=\textwidth]{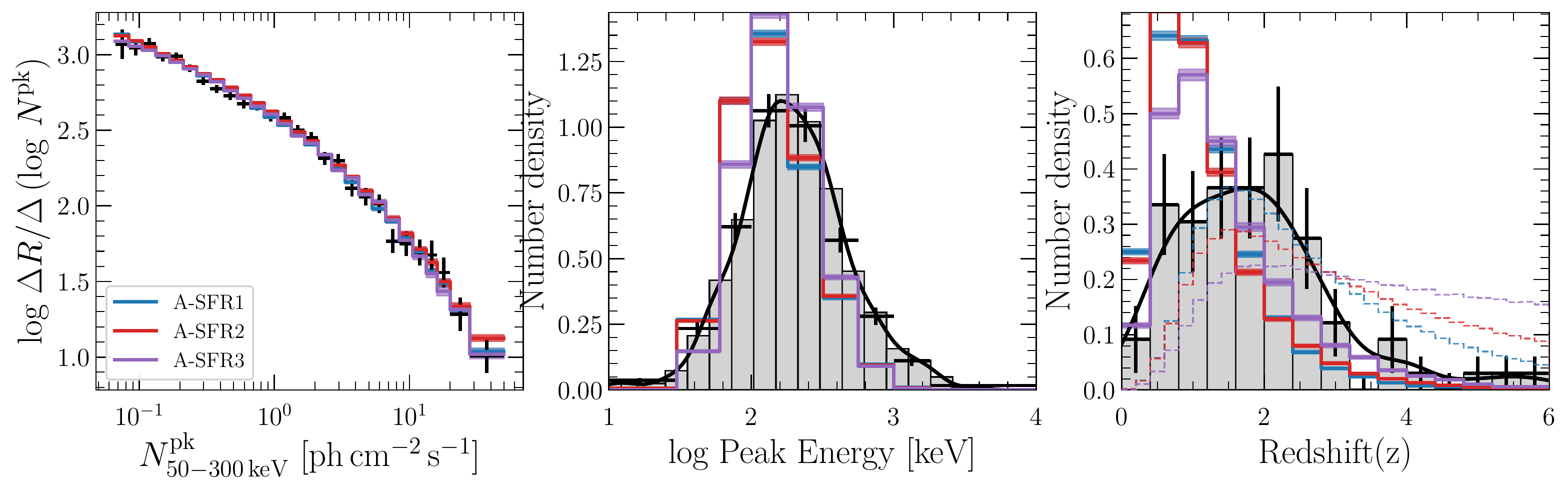}
\includegraphics[width=\textwidth]{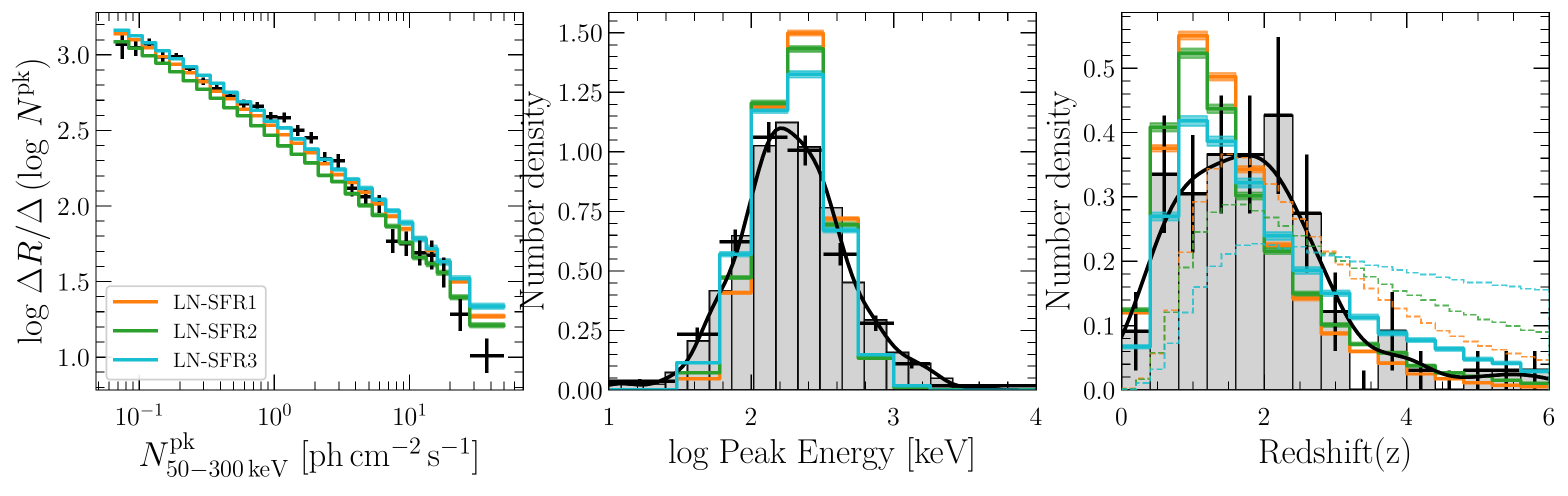}
\end{center}
\caption{Population generated using the model parameters of \citet{Daigne2006} compared to the observational constraints presented in Sect.~\ref{sec:obs_constraints}.
In the rightmost panels, the dashed lines correspond to the redshift distribution of the entire intrinsic population.
The top panels correspond to the intrinsic \EpL\ correlation scenarios and the bottom panels to the independent log-Normal \Ep\ scenarios.}
\label{fig:constraints_D06}
\end{figure*}

We used the best fit values presented in Table~1 of \citet{Daigne2006} for the six scenarios they explored which cover three different hypotheses concerning the LGRB efficiency (constant, mildly increasing and strongly increasing, labelled SFR1, SFR2, SFR3 respectively) and two scenarios for the peak energy distribution (intrinsically correlated with the peak luminosity, labelled A for "Amati-like", and independent log-normal, labelled LN).
We used a Band function for the LGRB spectra with their empirical distributions for $\alpha$ and $\beta$.
We used the same cosmology: $\Omega_{\rm m} = 0.3$, $\Omega_{\Lambda} = 0.7$, and $H_{0} = 65$~km~s$^{-1}$~Mpc$^{-1}$.
The observational constraints presented in Sect.~\ref{sec:obs_constraints} and the predictions from the model parameters are shown in Fig.~\ref{fig:constraints_D06}.
It should be noted that our intensity constraint (leftmost panels in Fig.~\ref{fig:constraints_D06}) is the same as the one they used to adjust their population parameters, however the other constraints are new.
This comparison illustrates how using a redshift constraint, which was not possible in 2006,  provides a clear advantage to better constrain the underlying intrinsic LGRB population.

\subsection*{Wanderman \& Piran 2010}
\begin{figure*}
\begin{center}
\includegraphics[width=\textwidth]{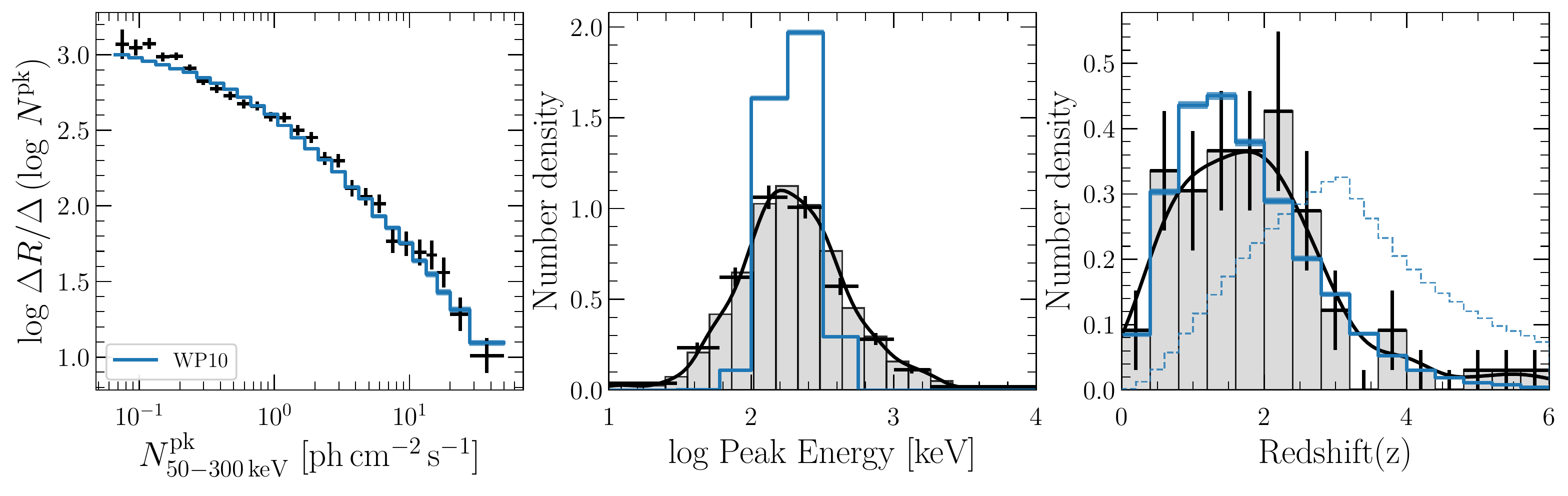}
\end{center}
\caption{Population generated using the model parameters of \citet{Wanderman2010} compared to the observational constraints presented in Sect.~\ref{sec:obs_constraints}; the intensity constraint was renormalised.}
\label{fig:constraints_WP10}
\end{figure*}

We used the best fit values reported in Table~1 of \citet{Wanderman2010} for the 1/2 bins case.
We used a broken power law for the luminosity function, using their value $\Lmin=10^{50}~\ergs$ and a fixed Band spectrum with $\Ep=511~\keV$, $\alpha=-1$ and $\beta=-2.25$.
We used the same cosmology: $\Omega_{\rm m} = 0.27$, $\Omega_{\Lambda} = 0.73$, and $H_{0} = 70$~km~s$^{-1}$~Mpc$^{-1}$.
The observational constraints presented in Sect.~\ref{sec:obs_constraints} and the predictions from the model parameters are shown in Fig.~\ref{fig:constraints_WP10}.
It should be noted their population is adjusted on a sample of $\sim120$ \textit{Swift} bursts, however it reproduces very nicely the intensity constraint of observed \BATSE\ LGRBs if the normalisation is adjusted.
The spectrum constraint is poorly reproduced because they used a fixed spectrum for all LGRBs, on the other hand the redshift constraint is surprisingly well reproduced which suggests that their approximate criteria for the probability of redshift measurement as a function of gamma-ray peak flux is reasonable.

\subsection*{Salvaterra et al. 2012}
\begin{figure*}
\begin{center}
\includegraphics[width=\textwidth]{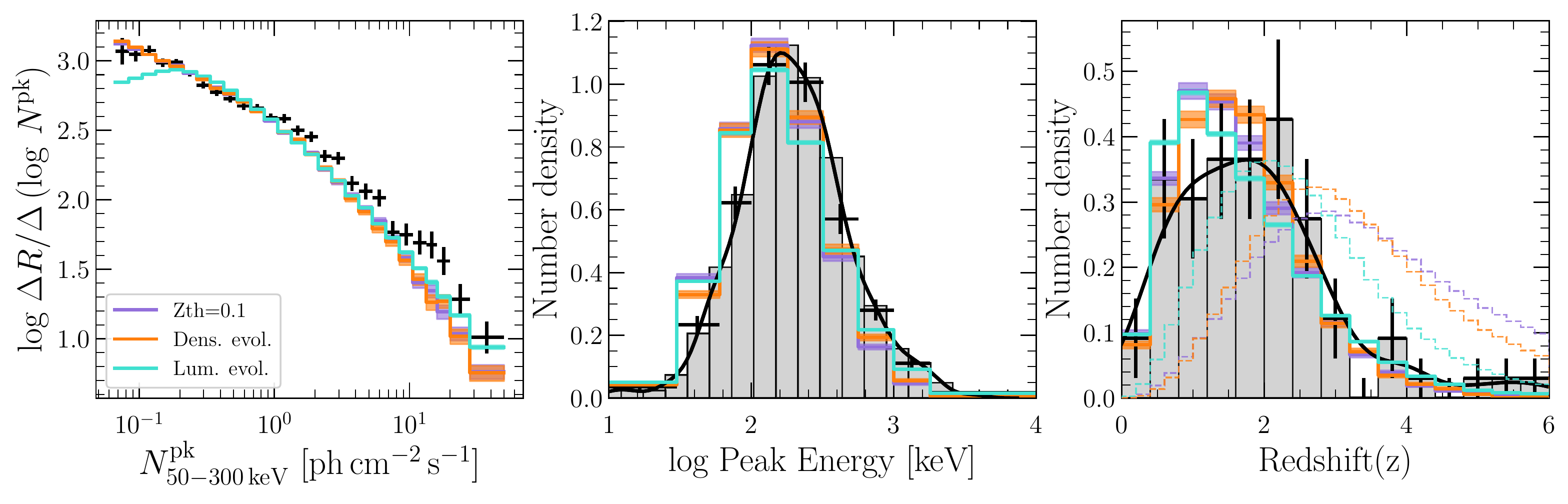}
\end{center}
\caption{Model parameters of \citet{Salvaterra2012} applied to the observational constraints presented in Sect.~\ref{sec:obs_constraints}.}
\label{fig:constraints_S12}
\end{figure*}

We used the best fit values presented in Table~2 of \citet{Salvaterra2012} for the luminosity evolution scenario, the density evolution scenario and the metallicity threshold scenario.
In all three scenarios we used the values for a broken power law luminosity function with $\Lmin=10^{49}~\ergs$ and a Band spectrum with $\alpha=-1$, $\beta=-2.25$ and with \Ep\ following the intrinsic correlation of Eq.~\ref{eq:intr_correl} with $\alphaA=0.49$, $\Epo=337~\keV$, $\sigmaEp=028$.
We used the same cosmology: $\Omega_{\rm m} = 0.3$, $\Omega_{\Lambda} = 0.7$, and $H_{0} = 70$~km~s$^{-1}$~Mpc$^{-1}$.
It should be noted that the luminosity evolution scenario of \citet{Salvaterra2012} is slightly different in its definition than what we used in our model.
In their implementation, it is $L_{\rm cut}$, the break in the power law, that evolves with redshift and \Lmin\ stays fixed, whereas in our implementation defined in Eq.~\ref{eq:lum_evol}, the whole luminosity function evolves with redshift, including \Lmin\ and \Lmax.
We therefore modified our code accordingly in order to properly reproduce their luminosity function and used the value of $\kevol=2.1$ for the luminosity evolution scenario; for the density evolution scenario and for the metallicity threshold scenario (see the description and the definition of parameters in \citealt{Salvaterra2012}) we used $\delta_n=1.7$ and $\rm Z_{\rm th}=0.1$ respectively.
The observational constraints presented in Sect.~\ref{sec:obs_constraints} and the predictions from the model parameters are shown in Fig.~\ref{fig:constraints_S12}.
In general, this population adjusted on the original \textit{Swift}/BAT6 sample of 58 LGRBs and the $\log{N}$-$\log{P}$ of \citet{Stern2001} reproduces well our observational constraints, despite some discrepancies at high and low peak flux.
The simulated peak energy distribution is also lower on average than the one observed by \GBM.
This illustrates the strength of simultaneously using constraints  from \BATSE, \GBM\ and \Swift\ that cover all major aspects of a population.
We extend this approach in the present paper adding a spectral constraint based on \GBM\ and using Swift data accumulated on a much much longer duration.

\subsection*{Pescalli et al. 2016}
\begin{figure*}
\begin{center}
\includegraphics[width=\textwidth]{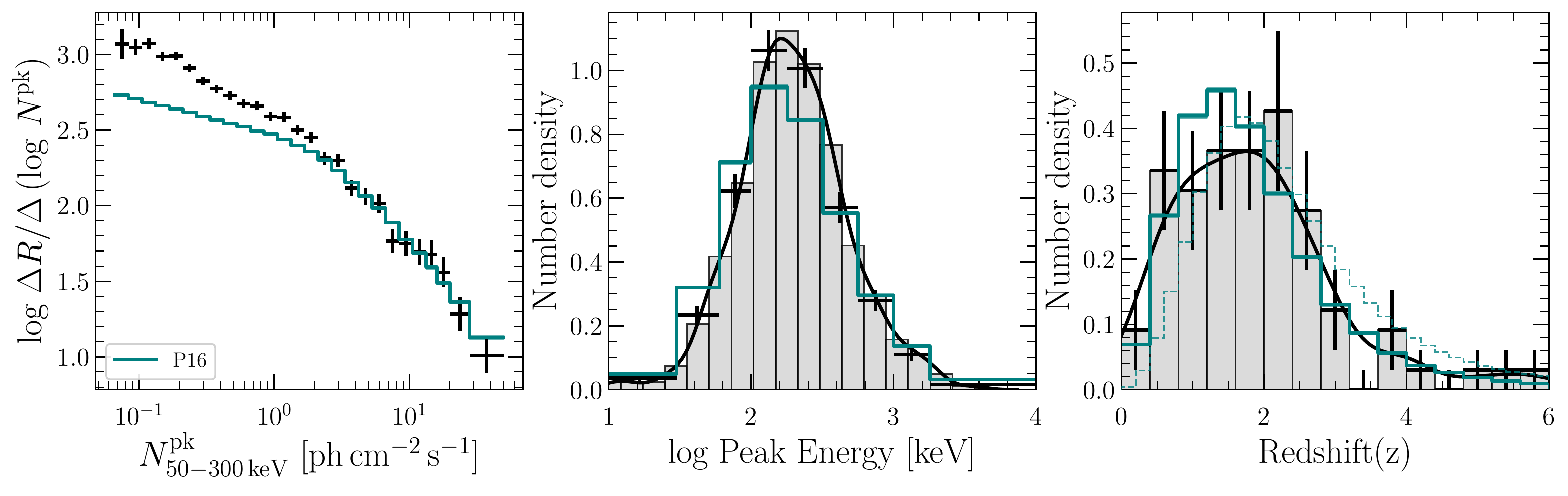}
\end{center}
\caption{Model parameters of \citet{Pescalli2016} applied to the observational constraints presented in Sect.~\ref{sec:obs_constraints}; the intensity constraint was renormalised.}
\label{fig:constraints_P16}
\end{figure*}

\citet{Pescalli2016} derived step functions for the LGRB formation rate and luminosity function, using a different approach than the previous studies cited above, based on the $C^-$ method of \citet{LyndenBell1971}.
In this work, contrary to \citet{Salvaterra2012}, the authors constraints rely only on the luminosity and redshift distributions (and their join distribution: the luminosity-redshift plane) of the extended BAT6 sample of bright \textit{Swift} LGRBs with $\Npksub{15}{150}{keV}~\geq~2.6~\phscm$.
We use the values they quote in Section~7 for a broken power law fit to their luminosity step function with $\Lmin=10^{49}~\ergs$ and $\kevol=2.5$.
They do not quote any analytical form for the redshift distribution but they conclude it follows the star formation rate; we therefore used the form of \citet{Madau2014}.
We used a Band spectrum with $\alpha=-1$, $\beta=-2.25$ and with \Ep\ following the intrinsic correlation of Eq.~\ref{eq:intr_correl} with the values they derive in Table~1: $\alphaA=0.54$, $\Epo=309~\keV$, $\sigmaEp=028$.
We used the same cosmology: $\Omega_{\rm m} = 0.3$, $\Omega_{\Lambda} = 0.7$, and $H_{0} = 70$~km~s$^{-1}$~Mpc$^{-1}$.
The observational constraints presented in Sect.~\ref{sec:obs_constraints} and the predictions from the model parameters are shown in Fig.~\ref{fig:constraints_P16}.
Interestingly, we see in our intensity constraint that this population reproduces well the high peak flux regime (after renormalisation) but departs from the $\log{N}$-$\log{P}$ towards lower peak fluxes.
This comes as a surprise since the luminosity functions of \citet{Salvaterra2012} and \citet{Pescalli2016} appear consistent in the right panel of Figure~1 of \citet{Pescalli2016}.
This is probably due to the difference in the low-luminosity slope between the two cases, which affects the number of faint bursts and thus the faint end of the $\log{N}$-$\log{P}$.
The discrepancies in the $\log{N}$-$\log{P}$ underline the advantage of constraining a population model on multiple complementary observational constraints that reach down to faint peak fluxes.

\end{document}